A. SOPCZAK
on behalf of the ATLAS Forward Detector group,
Institute of Experimental and Applied Physics, Czech Technical University in Prague, and
Ulrich Bonse Visiting Chair for Instrumentation, TU Dortmund University, Dortmund, Germany
*(Husova 240/5, CZ-11000 Prague; e-mail: andre.sopczak@cern.ch)*

# Overview of ATLAS forward proton detectors: status, performance and new physics results

*A key focus of the physics program at the LHC is the study of head-on proton-proton collisions. However, an important class of physics can be studied for cases where the protons narrowly miss one another and remain intact. In such cases, the electromagnetic fields surrounding the protons can interact producing high-energy photon-photon collisions. Alternatively, interactions mediated by the strong force can also result in intact forward scattered protons, providing probes of quantum chromodynamics (QCD). In order to aid identification and provide unique information about these rare interactions, the instrumentation to detect and measure protons scattered through very small angles is installed in the beam pipe far downstream of the interaction point. We describe the ATLAS Forward Proton 'Roman Pot' detectors (AFP and ALFA), their performance of Tracking and Time-of-Flight detectors, and first results.*

*Keywords*: ATLAS Forward Proton detectors, AFP, ALFA

## 1. Introduction

The ATLAS Forward Proton (AFP) project significantly extends the ATLAS physics program by tagging and measuring the momentum and emission angle of very forward protons[1]. This enables the observation and measurement of a range of processes where one or both protons remains intact which otherwise would be difficult or impossible to study. Such processes are typically associated with elastic and diffractive scattering, where the proton radiates either a photon or a virtual colorless object, the so-called Pomeron, which is often thought of as a non-perturbative collection of soft gluons. The article is structured as follows:

- Physics motivation
- ATLAS [1] Roman Pots
- ALFA detector
- Total cross-section measurement
- Measurement of exclusive pion pair production
- AFP detector
- SiT detector
- LHC Run-3 data-taking
- Data Quality (DQ)
- SiT hit map
- SiT track map
- Correlation AFP and ATLAS central detectors
- ToF-SiT alignment
- ToF efficiency
- ToF vertex matching

[1] This work is based on the results presented at the 2024 "New Trends in High-Energy and Low-x Physics" Conference.

- ToF performance in LHC Run-2
- AFP results
- Matching of proton energy loss with ATLAS central di-leptons/di-photons events
- Di-lepton production in photon collisions
- Axion-Light-Particle (ALP) search in Light-by-Light scattering
- Comparison with previous ALP results and extrapolations

## 2. Physics Motivation

Usually, in proton-proton collisions at the LHC, the proton breaks up, as shown in Fig. 1. However, in proton-proton interactions via a photon ($\gamma$) exchange, electromagnetic force, or pomeron (P) exchange, strong force, the proton can remain intact (Fig 2).

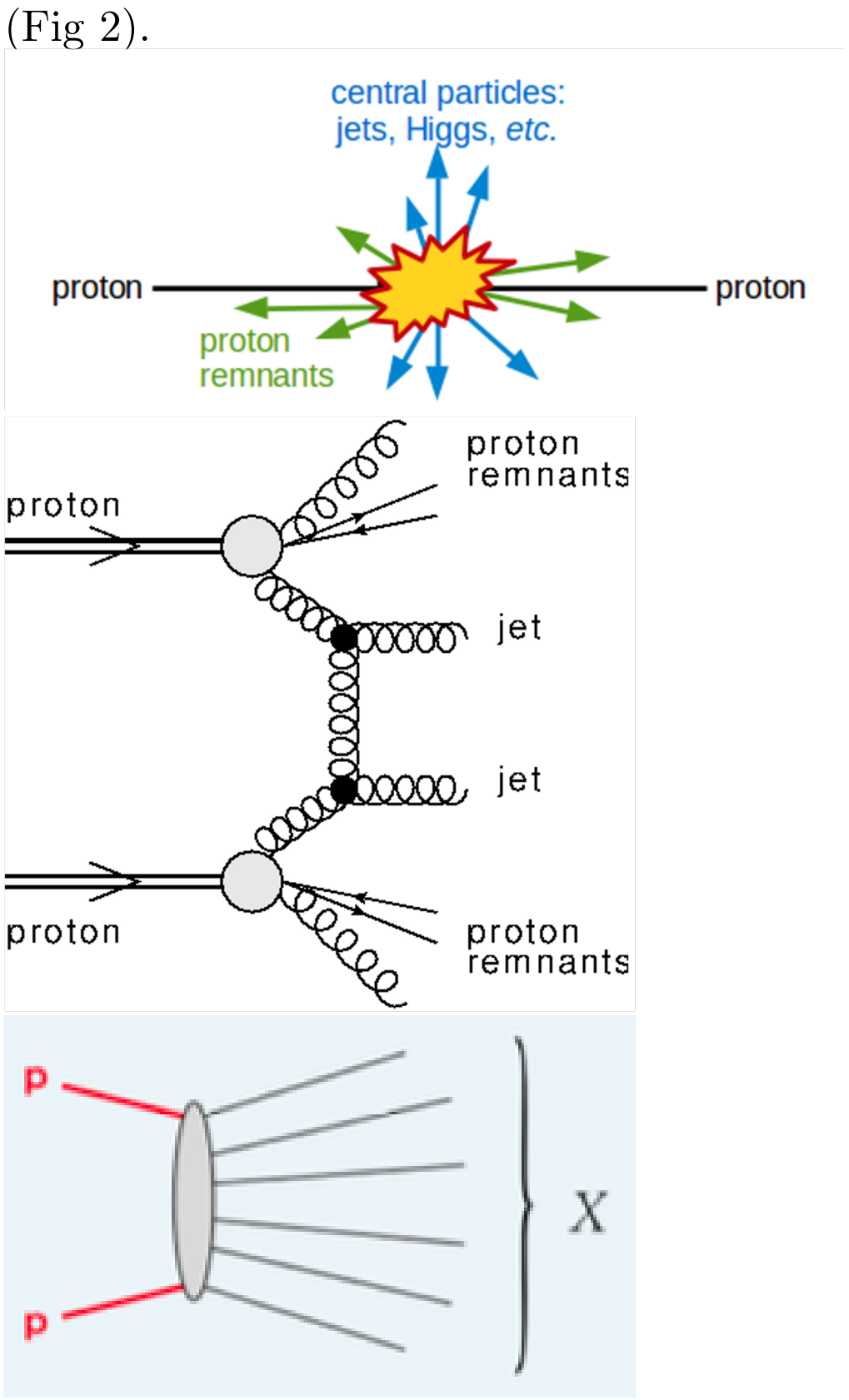

***Fig. 1.*** Typical proton-proton collisions.

The detection of events containing scattered intact protons focuses on low-cross section processes with high pT objects. Examples are given in Fig. 3.

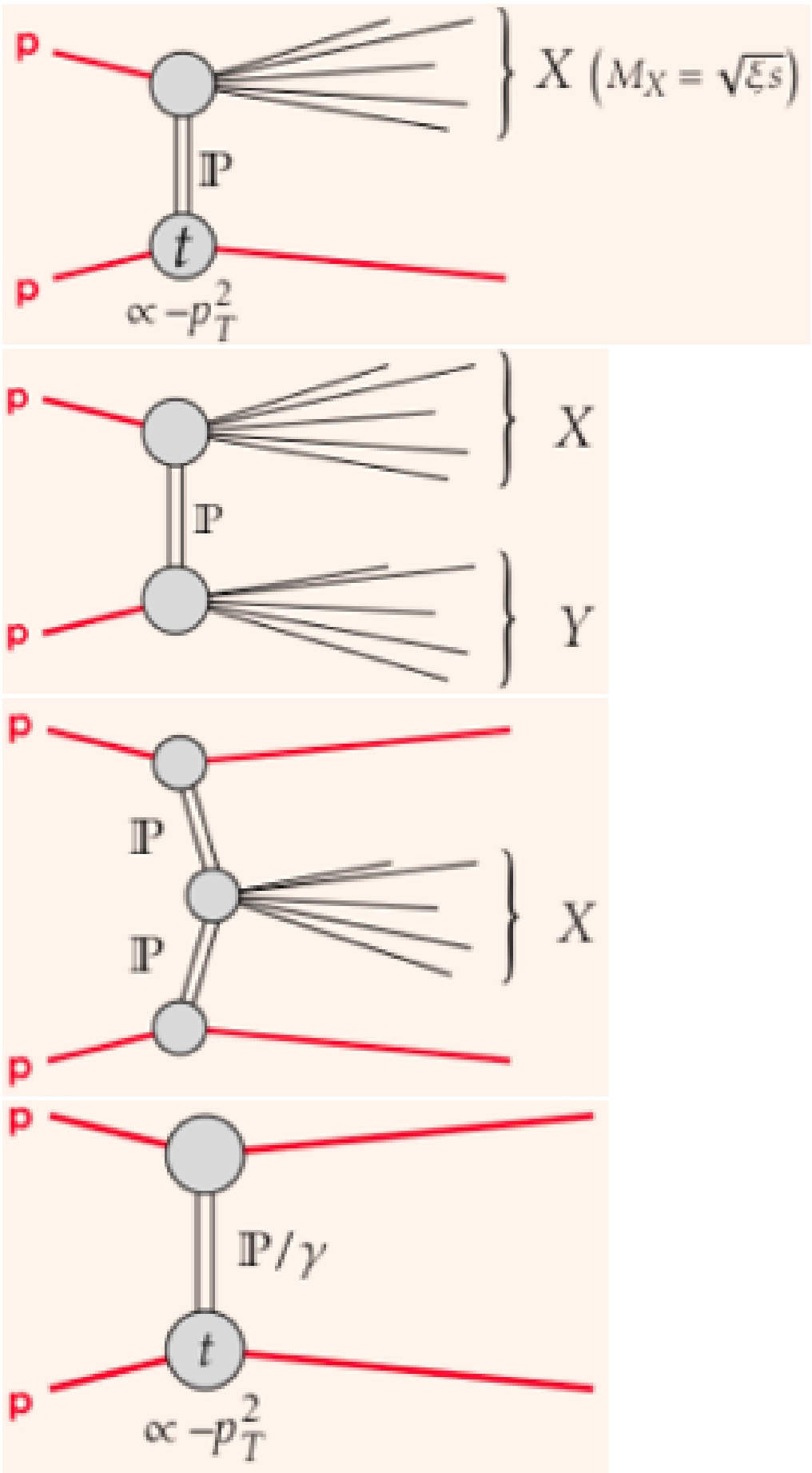

***Fig. 2.*** Single diffractive, double diffractive, central diffractive and elastic scattering.

## 3. ATLAS Roman Pots

The forward detectors are located in the LHC tunnel outside the ATLAS cavern. During a physics run, they are moved close to the beam (1-3 mm) once "Stable Beams" are declared. There are two sub-detector systems:

- Absolute Luminosity For ATLAS (ALFA), 237 m and 245 m from the interaction point (IP).
- ATLAS Forward Proton (AFP), 205 m and 217 m from the IP.

Figure 4 [2] shows a schematic overview of the ALFA and the AFP locations in the LHC tunnel.

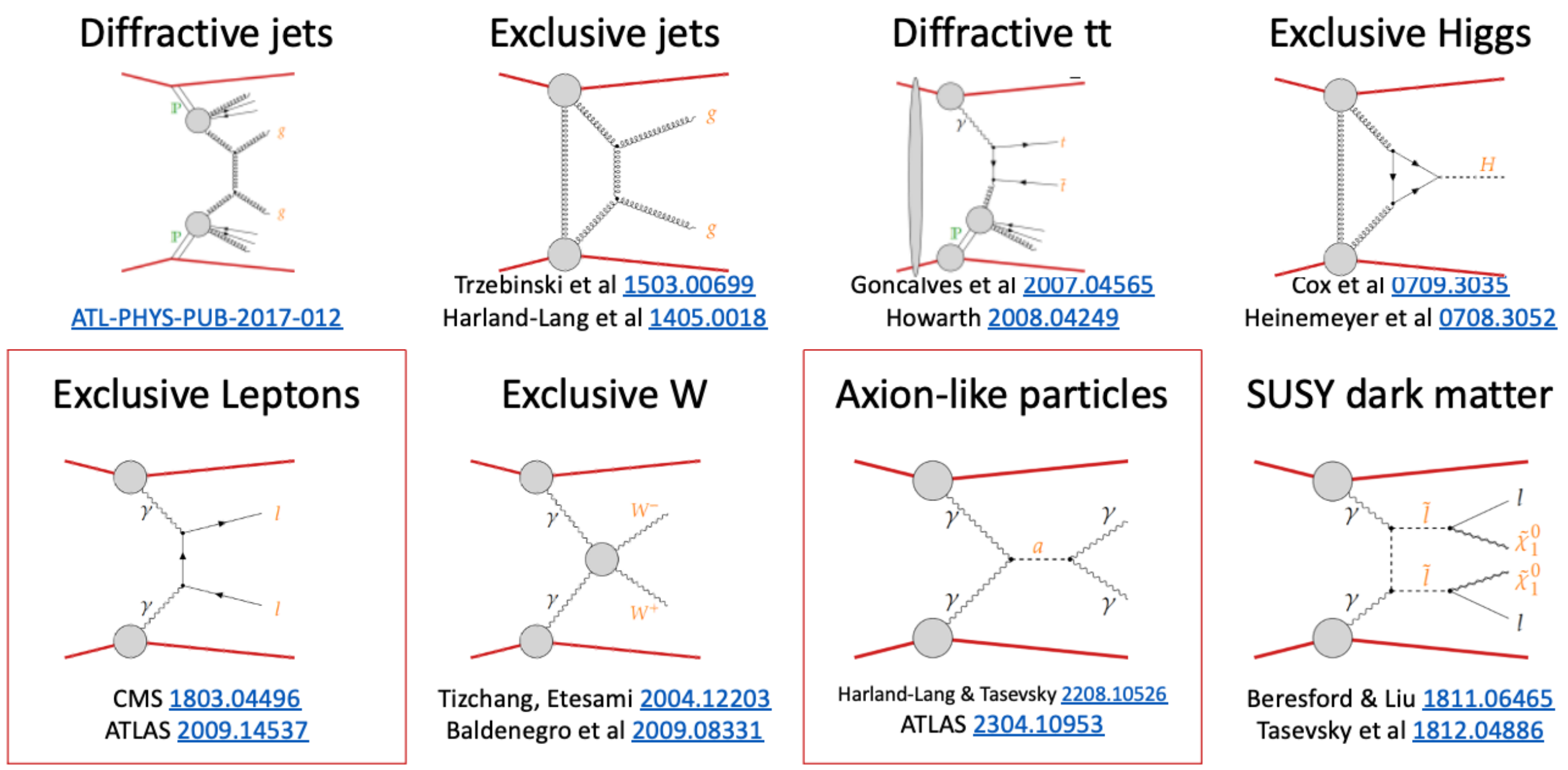

***Fig. 3.*** Processes containing intact protons. The two processes in the red boxes have recently been analysed by the ATLAS collaboration.

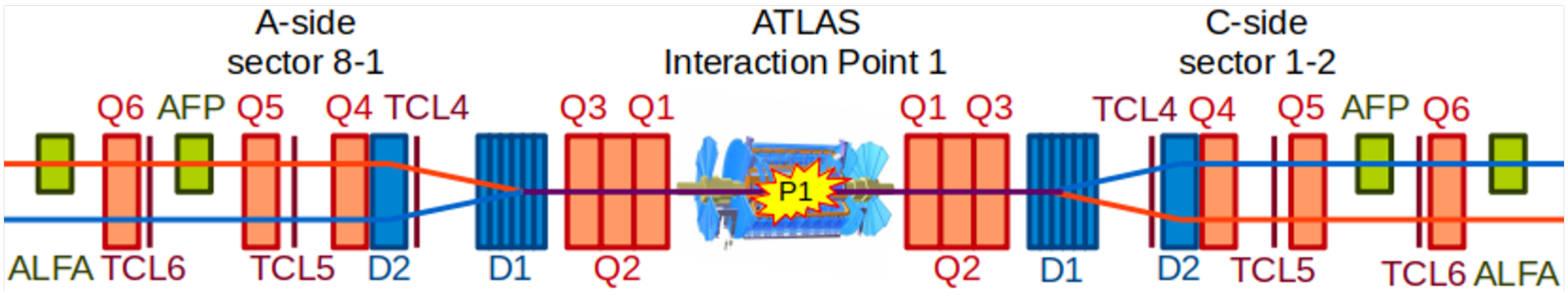

***Fig. 4.*** Overview of the ALFA and the AFP locations in the LHC tunnel (from [2]).

## 4. ALFA Detector

The ALFA detector is a specific part of the ATLAS experiment designed to measure the elastic scattering of protons. It has two stations with tracking detectors located on both sides of the central ATLAS detector. The reconstruction efficiency is measured by a tag-and-probe method (well-measured protons on one side as tags for a proton on the other side). Important systematics are the reconstruction efficiency uncertainty of 0.4% - 0.9%, dominated by the evaluation of accidental coincidences and uncertainties in backgrounds [2]. The tracking accuracy is dominated by the global vertical distance uncertainty (after alignment) of $\pm 22$ microns. Figure 5 gives an overview of the ALFA detector and shows the reconstruction efficiency during the $\beta^* = 2.5$ km data-taking campaign [2].

The ALFA data-taking is summarized in Fig. 6. ALFA did not run with high LHC luminosity, because the detector is radiation-sensitive. In standard high-luminosity LHC running, beams are focused to a small region at the ATLAS interaction point. Protons at different angles were focused together and emerge in a broad beam. ALFA running needed to measure pp to pp elastic cross sections down to low scattering angles $\theta$. Outgoing protons at different $\theta$ were detected at different positions $y$ at ALFA. Thus, "parallel to point" vertical focusing from the IP requires large values of the beam parameter $\beta^*$ at the IP. This implies larger beam size at the IP and, therefore low pile-up is needed for these measurements.

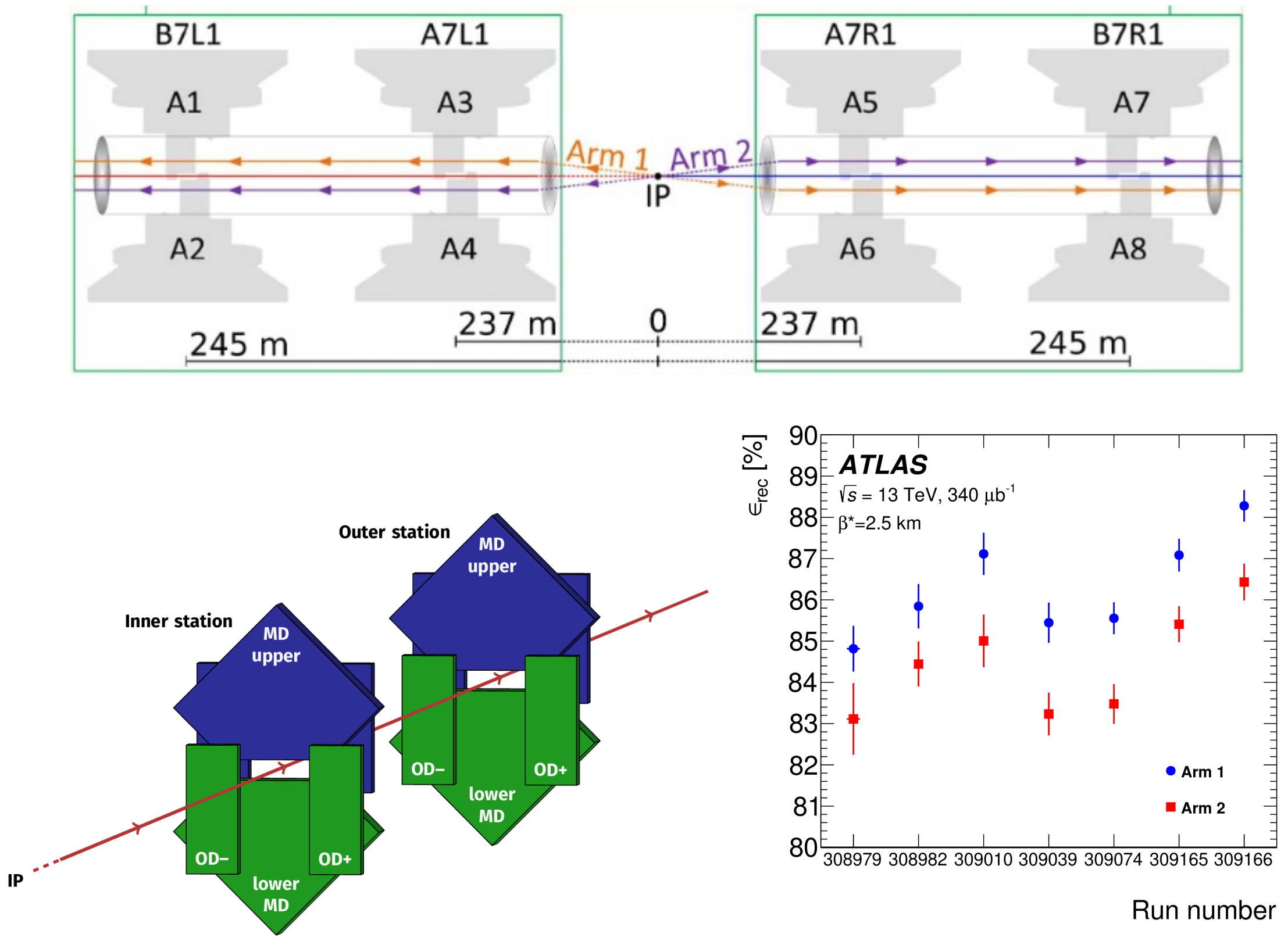

***Fig. 5.*** Schematic view of the ALFA detector and the reconstruction efficiency (from [2]).

| Year | β* | √s [TeV] | Comments |
|---|---|---|---|
| 2011 | 90 m | 7 | elastics: NPB 889 (2014)<br>excl. $\pi^+\pi^-$: EPJC 83 (2023) 627 |
| 2012 | 90 m | 8 | elastics: PLB 761 (2016)<br>single diff.: JHEP 02 (2020) 042 |
| 2012 | 1 km | 8 | elastics dataset |
| 2013 | 0.8 m | 2.76 | proton-lead dataset |
| 2013 | 0.8 m | 2.76 | proton-proton reference dataset |
| 2015 | 90 | 13 | diffractive dataset |
| 2016 | 2.5 km | 13 | elastics: EPJC 83 (2023) 441 |
| 2018 | 90 m | 13 | elastic (large *t*) and diff. datasets |
| 2018 | 11 m | 0.9 | elastics (large *t*) dataset |
| 2018 | 50/100m | 0.9 | elastics dataset |
| 2023 | 3/6 km | 13.6 | elastics dataset |

Measurement of the total cross section from elastic scattering in pp collisions at √s = 7 TeV with the ATLAS detector, Nucl. Phys. B (2014) 486. $\sigma_{tot}$ = 95.35 ± 1.36 mb
One dedicated run at at β* = 90 m, integrated luminosity 80 μb$^{-1}$
Measurement of the total cross section from elastic scattering in pp collisions at √s = 8 TeV with the ATLAS detector, Phys. Lett. B 761 (2016) 158. $\sigma_{tot}$ = 96.07 ± 0.92 mb
One dedicated run at at β* = 90 m, integrated luminosity 500 μb$^{-1}$
Measurement of differential cross sections for single diffractive dissociation in √s = 8 TeV pp collisions using the ATLAS ALFA spectrometer, JHEP 2020 (2020) 42.
One dedicated run at at β* = 90 m, integrated luminosity 500 μb$^{-1}$
Measurement of the total cross-section and ρ-parameter from elastic scattering in pp collisions at √s = 13 TeV with the ATLAS detector, Eur. Phys. J. C 83 (2023) 441.
Seven dedicated runs at β* = 2500 m, total integrated luminosity 340 μb$^{-1}$
Measurement of exclusive pion pair production in proton–proton collisions at √s = 7 TeV with the ATLAS detector. Eur. Phys. J. C 83 (2023) 627.
One dedicated run at at β* = 90 m, integrated luminosity 80 μb$^{-1}$.

***Fig. 6.*** ALFA data-taking and published results.

## 5. Total Cross Section Measurement

A few highlights from the $\sqrt{s} = 13\,\text{TeV}, \beta^* = 2.5\,\text{km}$ analysis are presented. Using ALFA data, the selection of elastic pp events is based on [2]:

- Quality cuts on the two proton tracks in the two ALFA stations
- Geometric acceptance cuts: Select back-to-back events, as indicated in Fig. 7 [2].
- Selection on $x$ versus $\theta_x$: elastic events are within the ellipse shown in Fig. 7 [2].

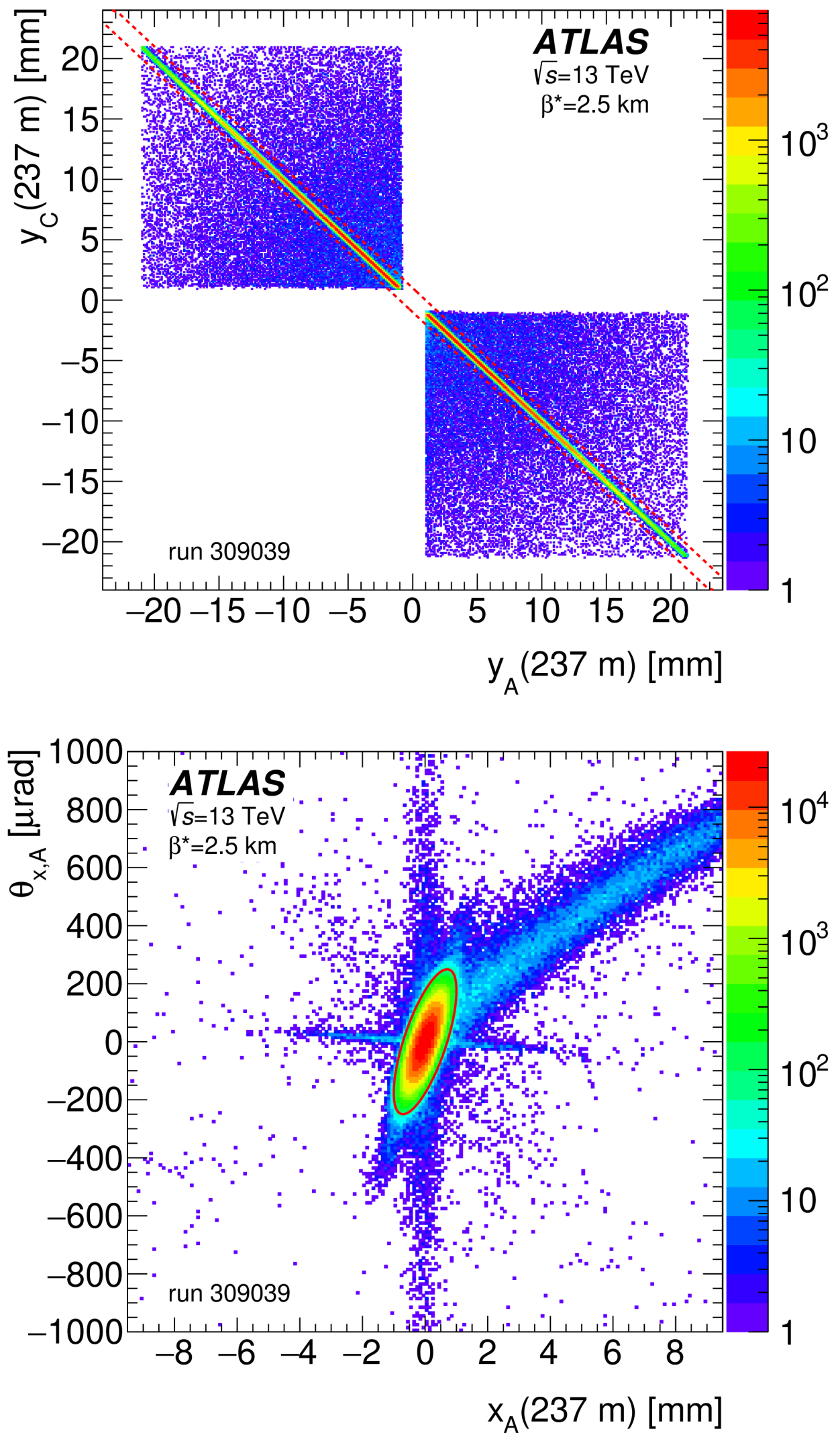

***Fig. 7.*** ALFA event selection (from [2]).

Sources of the background are:

- Accidental (beam) halo+halo and halo+soft single-diffractive proton coincidences (data-driven, determined with an event-mixing method).
- Central diffraction (MC simulation), double-Pomeron exchange (DPE), $pp \to pp + X$.

The Mandelstam variable $t$ is reconstructed from the beam optics and event kinematics using the tracking of effective beam optics. The $-t$ distribution is shown in Fig. 8 [2].

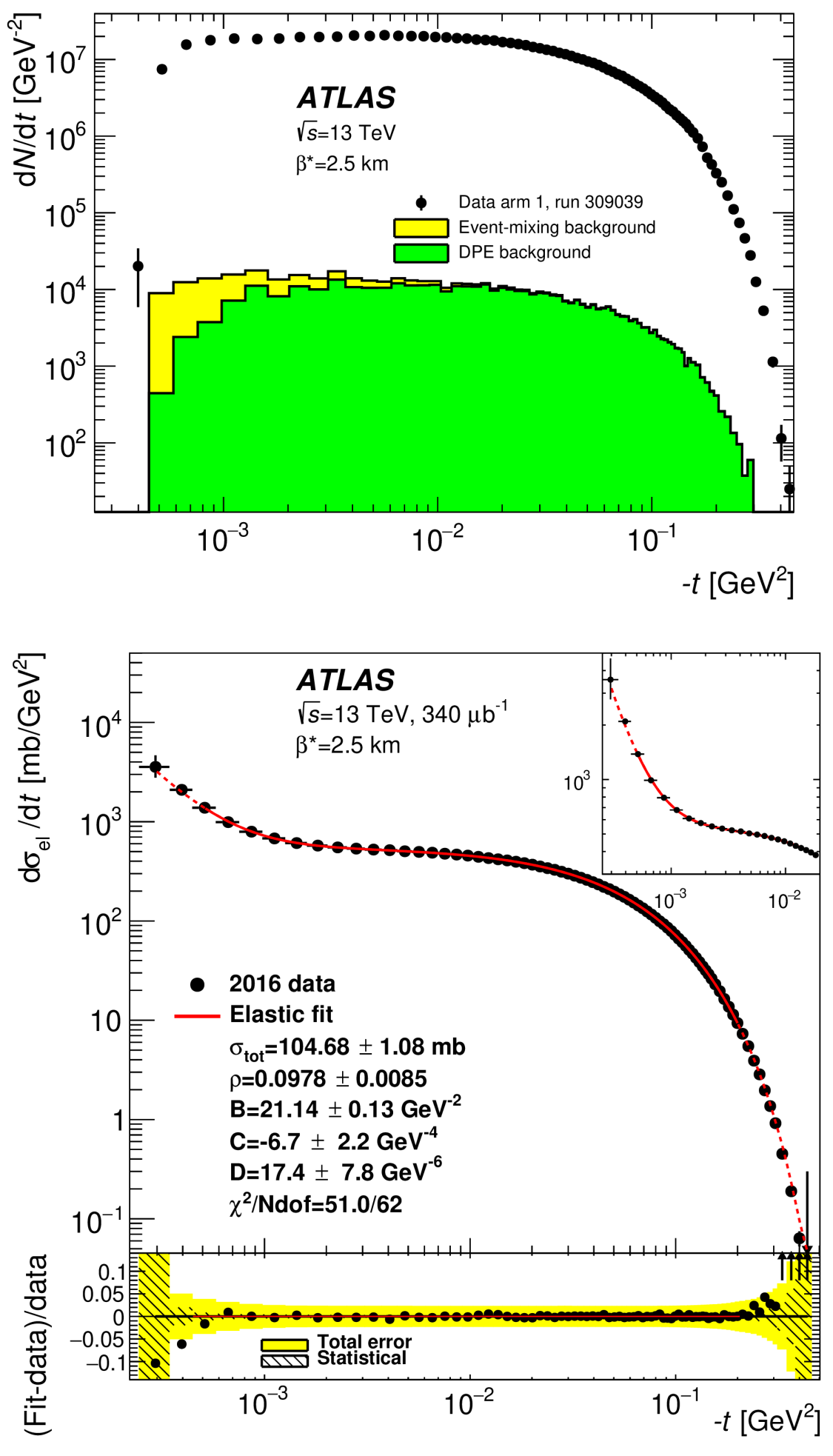

***Fig. 8.*** Mandelstam variable $-t$ distribution and fit (from [2]).

Elastic scattering is predominantly a low-$p_T$ process, and a perturbative expansion cannot be applied.

Therefore, $\sigma_{\text{tot}}$ and the $\rho$-parameter cannot be calculated from first principles in QCD. The $\rho$-parameter is defined as the ratio of the real part to the imaginary part of the elastic-scattering amplitude in the limit $t \to 0$. The results for $\sigma_{\text{tot}} = 104.68 \pm 1.08(\text{exp.}) \pm 0.12(\text{th.})$ mb and the $\rho = 0.098 \pm 0.011$ are shown in Fig. 9 in comparison with other measurements.

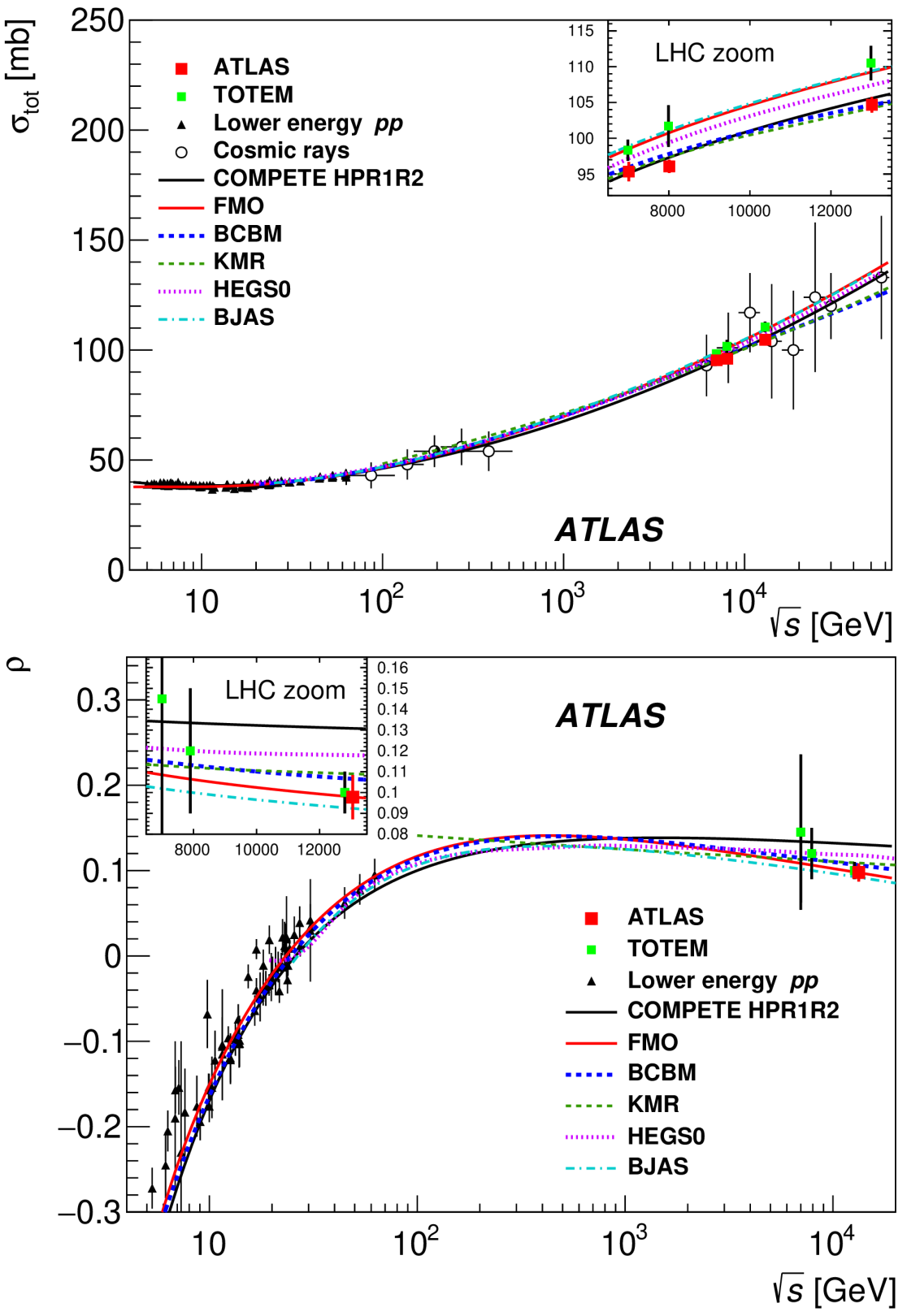

***Fig. 9.*** ATLAS $\sigma_{\text{tot}} = 104.68 \pm 1.08(\text{exp.}) \pm 0.12(\text{th.})$ mb and $\rho = 0.098 \pm 0.011$ measurements in comparison with other measurements (from [2]).

## 6. Measurement of Exclusive Pion Pair Production

The initial ALFA physics programme was extended towards diffractive studies. An example is the measurement of exclusive pion pair production in proton-proton collisions conducted at $\sqrt{s} = 7$ TeV with the ATLAS central and ALFA detectors [3]. The trigger selected:

- Elastic - ALFA coincidence of detectors in an elastic combination, and
- Anti-elastic - signal in any ALFA detector, prescaled by a factor 15.

In the ALFA detectors, one good quality track on each side is required, and in the ATLAS Inner Detector, two oppositely charged tracks, taken as pions, with $|\eta(\pi)| < 2.5$, $p_T(\pi) > 0.1$ GeV and quality requirements on the pion tracks are applied.

Further requirements are:

- Minimum-Bias Trigger Scintillators (MBTS) veto: at most one hit in the combined inner MBTS scintillators (at $z = \pm 3.6$ m, $2.1 < |\eta| < 3.8$), to remove diffractive-dissociative and non-diffractive events.
- Overall momentum balance: $pp\pi^+\pi^-$ momentum balance in $x$ and in $y$ consistent with zero ($\pm 3.5\sigma$).
- Track condition: tracks must have sufficient hits in MD layers, with a limit on the number of multiple hits in a layer.

Figure 10 shows the selected events [3].

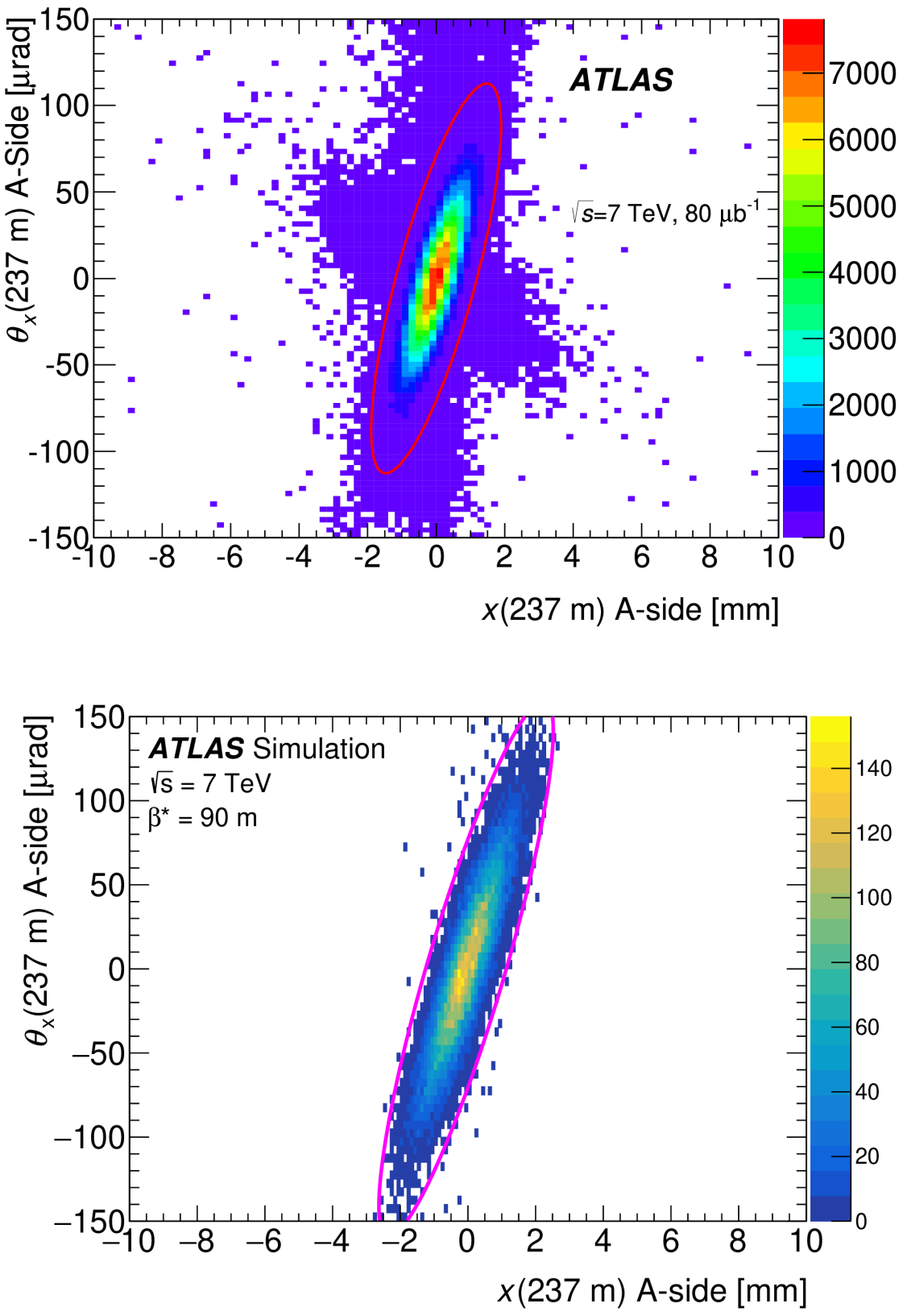

***Fig. 10.*** Selected events (red ellipse) in the ALFA detector. The ellipse is imposed on the ALFA tracks to reduce background events (from [3]).

The cuts are very effective at removing background events. The cut on the MBTS counts was essential. Low statistics from a short run in 2011 at 7 TeV (4 hours at high $\beta^*$, $\mu = 0.035$) was used. The feasibility of the measurement has been demonstrated. Figure 11 [3] shows signal and background events before and after the selection, and sources of systematic uncertainties. The uncertainty in the specification of the inner ATLAS detector material is the leading systematic uncertainty.

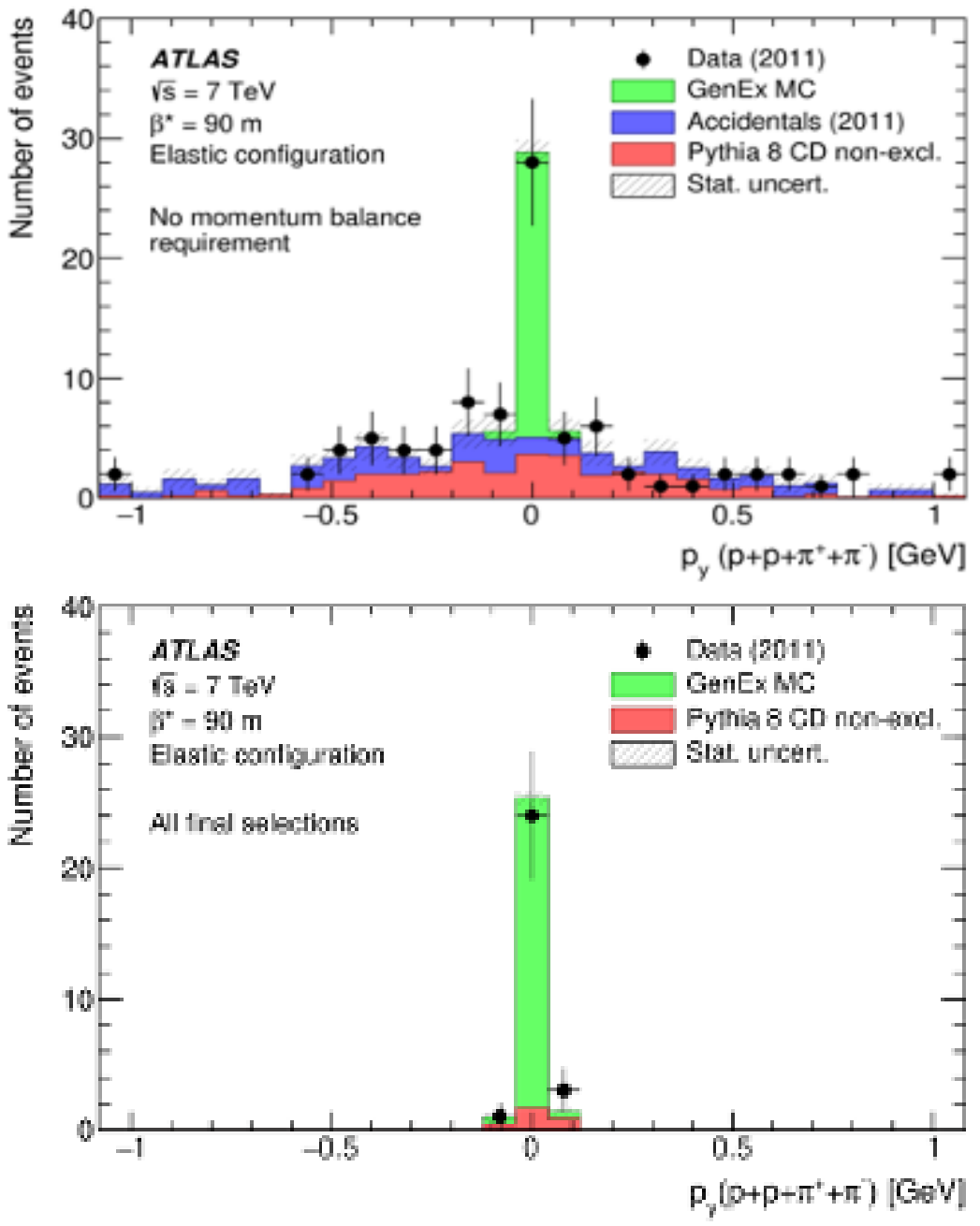

| Source of uncertainty | Uncertainty [%] elastic | anti-elastic |
|---|---|---|
| Trigger efficiency $\epsilon_{\rm trig}$ | ±0.1 | ±0.3 |
| Background determination | ±3.5 | ±3.5 |
| Signal and background corrections: | | |
| Beam energy | ±0.1 | ±0.1 |
| ID material | +4.8 | +4.1 |
| Veto on MBTS signal | ±1.3 | ±2.0 |
| ALFA single-track selection | ±0.9 | ±0.9 |
| ALFA reconstruction efficiency | ±0.9 | ±0.8 |
| ALFA geometry selection | ±0.5 | ±0.5 |
| Optics | ±1.1 | ±1.0 |
| Overall systematic uncertainty | $^{+6.4}_{-4.2}$ | $^{+6.0}_{-4.4}$ |
| Statistical uncertainty | ±21.2 | ±61.6 |
| Theoretical modelling | ±2.8 | ±8.0 |
| Luminosity | ±1.2 | ±1.2 |

***Fig. 11.*** Signal and background events before and after the selection, and source of systematic uncertainties (from [3]).

The outlook of ALFA detector is:

- From the final run with ALFA completed in LHC Run-3 (2023) using $\beta^* = 3/6$ km, an improved $\rho$ measurement for total cross section and parameters of elastic $pp$ scattering.
- Exclusive pion pair analysis using Run-2 dataset:
  - resonance analysis
  - possible glueball search
  - possible search for other exclusive final states
- Data at 900 GeV (2018) $\beta^* = 50/100$ m for total cross section and $\rho$ measurements.
- High-luminosity 0.5 nb$^{-1}$ at 13 TeV, $\beta^* = 90$ m for the study of dip/bump in $t$ distribution.

## 7. AFP Detector

The AFP detector has two stations on each side of the central ATLAS detector. All stations host Silicon Tracker (SiT) detectors, and far stations host also Time-of-Flight (ToF) detectors. Figure 12 shows the AFP detector layout and photos of the SiT and ToF detectors in the far stations.

## 8. SiT Detector

The main characteristics of the SiT detector are:

- The SiT detector is used for position measurement of scattered protons.
- The reconstruction of its kinematics uses 4 silicon pixel sensors, spaced 9 mm apart, each sensor has 336x80 pixels with a pixel size 50x250 $\mu$m$^2$, and sensor size 16.8x20 mm$^2$.
- Read out by FE-I4B chips, the same as for ATLAS Pixel IBL.
- Sensors have a 14° angle with respect to the beam axis to improve the track reconstruction resolution (about 6 $\mu$m in $x$ and about 30 $\mu$m in $y$).

## 9. LHC Run-3 Data-Taking

The total luminosity recorded in LHC Run-3 in 2022, 2023 and 2024 is 167.7 fb$^{-1}$ for the AFP. This is 91.6% with respect to ATLAS recorded and 86.0% with respect to LHC delivered [4].

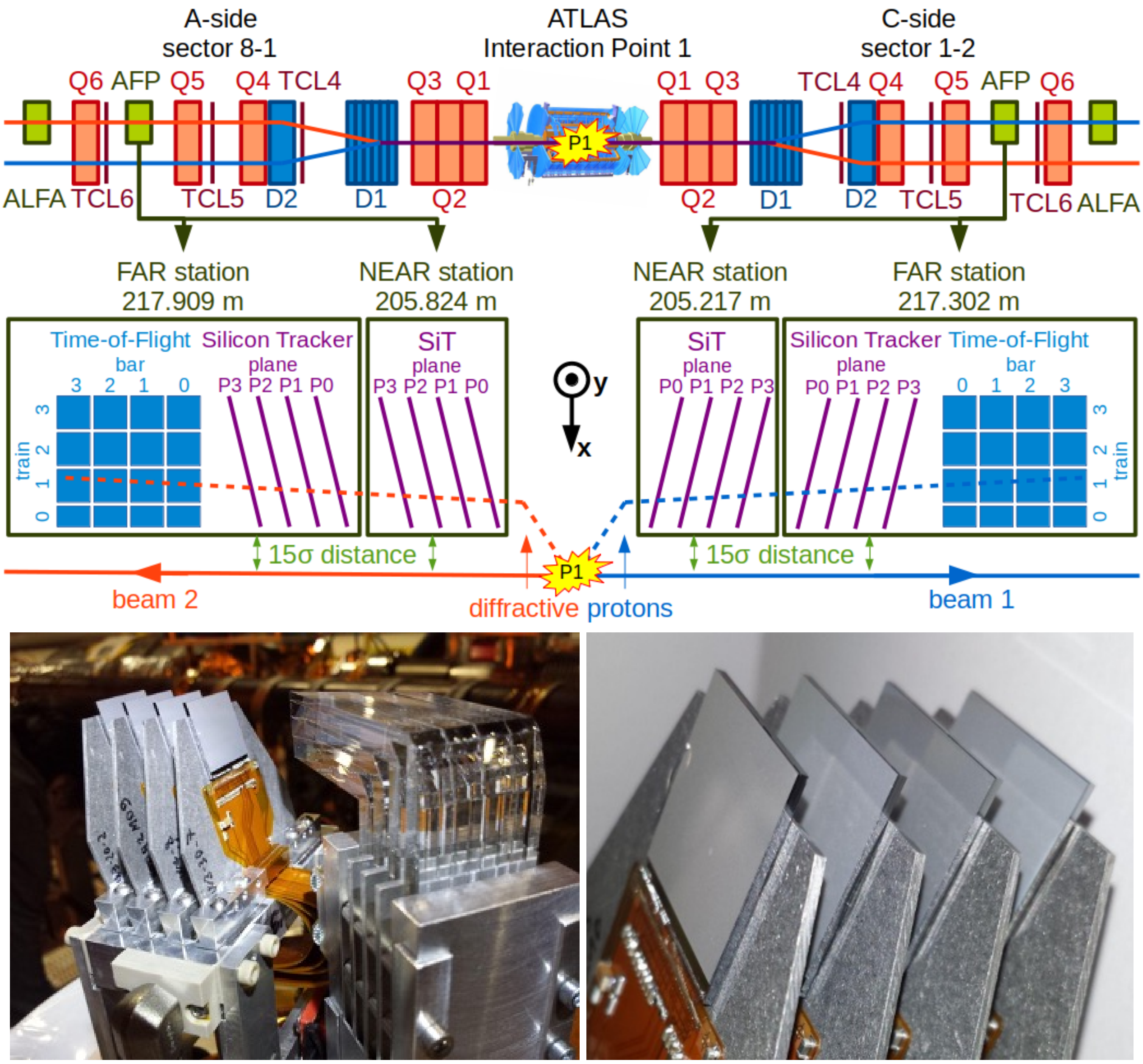

***Fig. 12.*** AFP detector layout and photos of the SiT and ToF detectors.

The preliminary integrated luminosities per year are:

- 2022 at $\sqrt{s}$ = 13.6 TeV, AFP recorded: 34.1 $fb^{-1}$, 95.5% with respect to ATLAS recorded, 88.6% with respect to LHC delivered.
- 2023 at $\sqrt{s}$ = 13.6 TeV, AFP recorded: 26.1 $fb^{-1}$, 87.9% with respect to ATLAS recorded, 82.3% with respect to LHC delivered.
- 2024 at $\sqrt{s}$ = 13.6 TeV, AFP recorded: 107.5 $fb^{-1}$, 91.4% with respect to ATLAS recorded, 86.2% with respect to LHC delivered.

Figures 13, 14 and 15 show the luminosity developments for 2022, 2023 and 2024 data-taking [4].

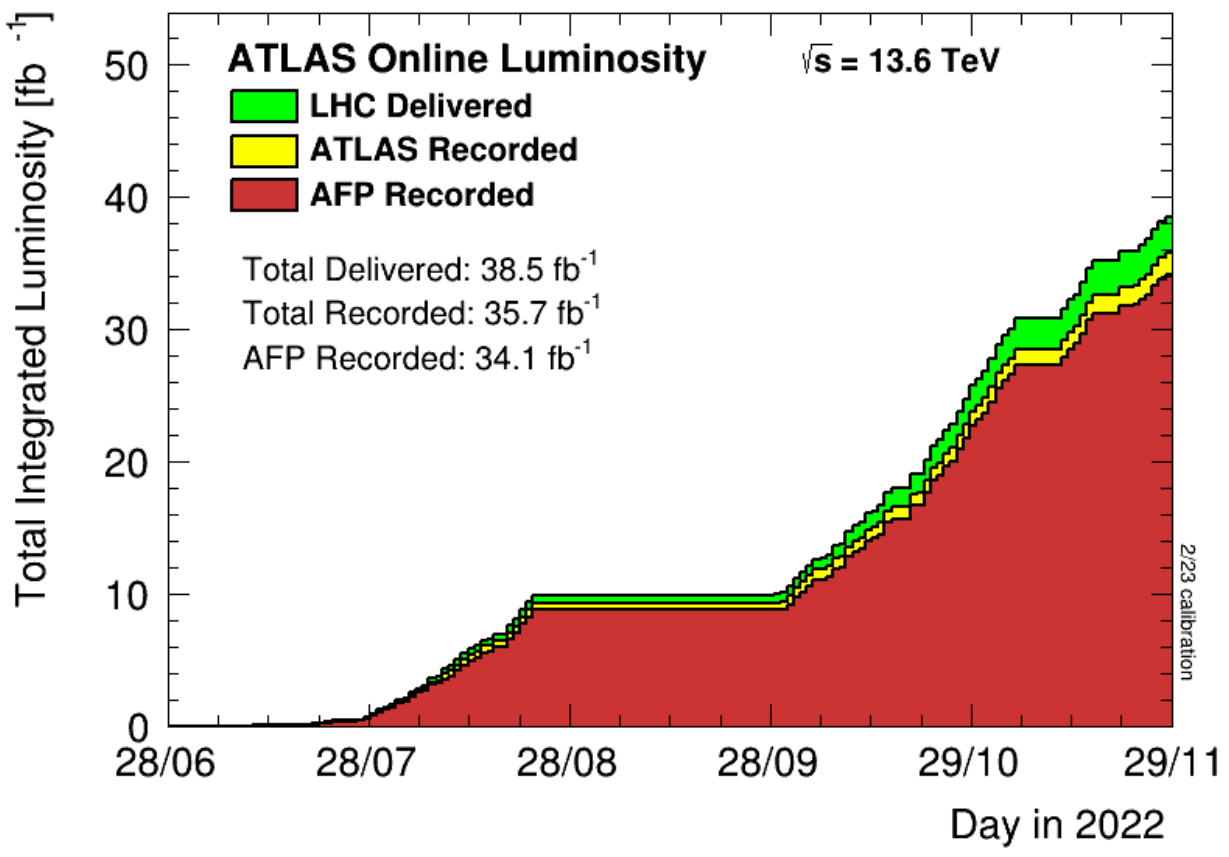

***Fig. 13.*** AFP luminosity development in 2022 (from [4]).

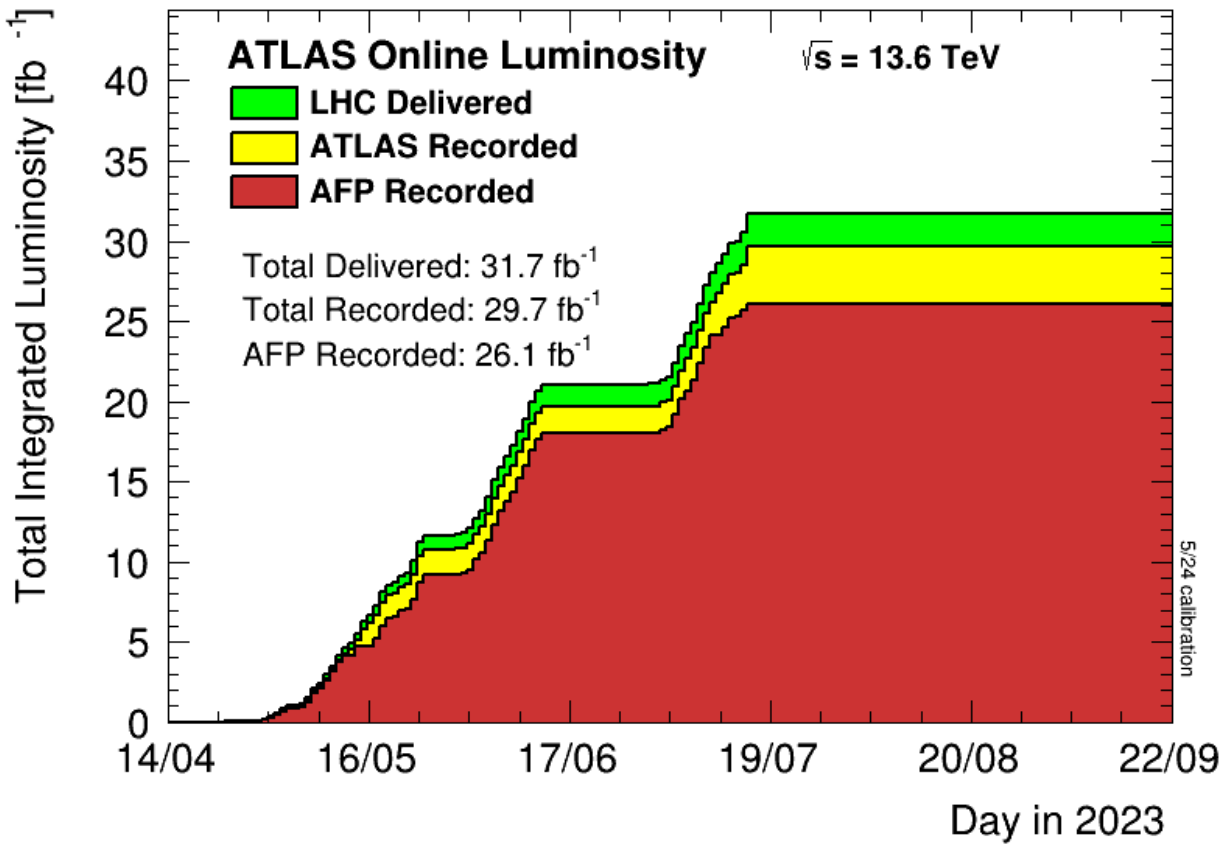

***Fig. 14.*** AFP luminosity development in 2023 (from [4]).

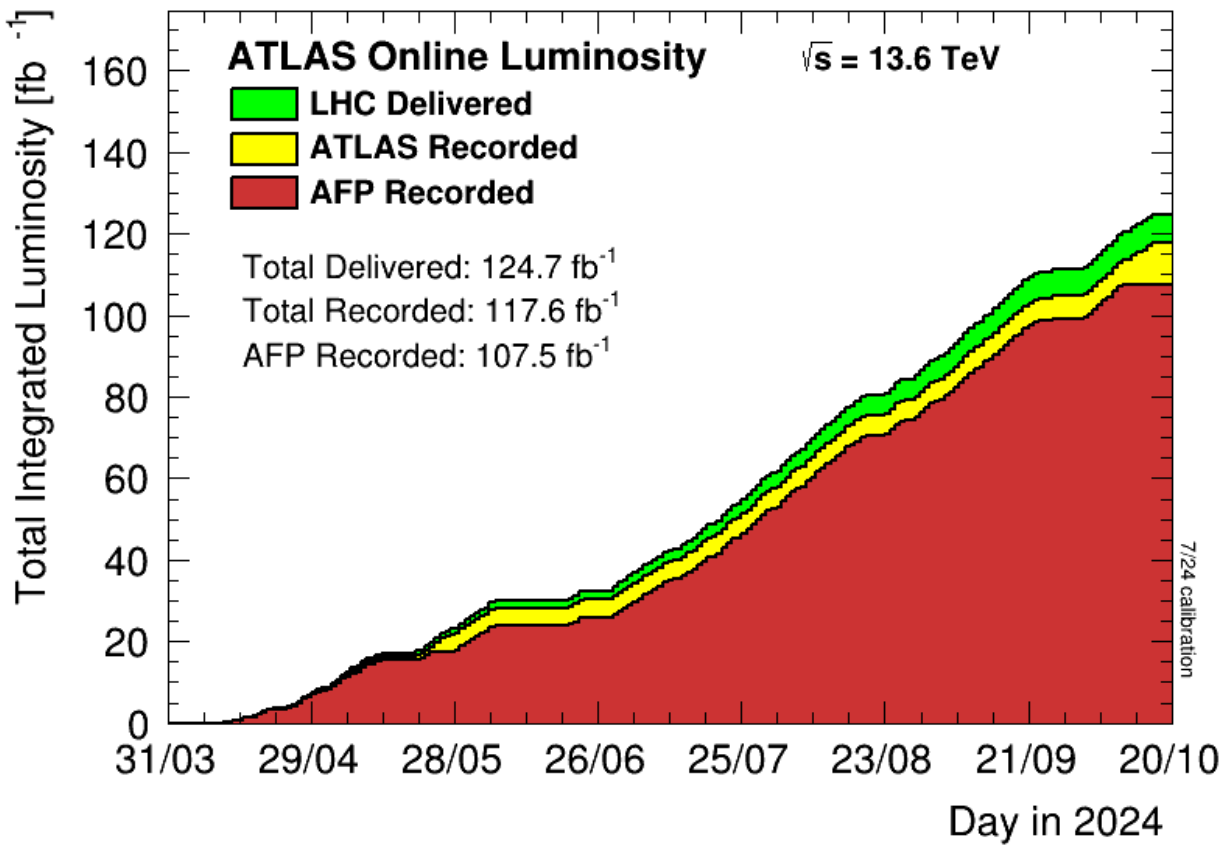

***Fig. 15.*** AFP luminosity development in 2024 (from [4]).

## 10. Data Quality (DQ)

The fraction of good luminosity after DQ criteria are applied with respect to the total ATLAS recorded luminosity. It is used as a measure of the DQ (Fig. 16) [3].

Fraction of good luminosity after Data Quality wrt. ATLAS:

| | 2022* | 2023** preliminary |
|---|---|---|
| All of AFP | 83.4 % | 76.4 % |
| Silicon Tracker only | 92.5 % | 81.4 % |
| A side Silicon Tracker only | 96.8 % | 84.5 % |
| C side Silicon Tracker only | 93.7 % | 82.1 % |
| Time-of-Flight only | 83.6 % | 77.7 % |

*based on Good Run List for analyses relying on jet, met or b-jet triggers (data22_13p6TeV.periodAllYear_DetStatus-v109-pro28-04_MERGED_PHYS_StandardGRL_All_Good_25ns)

**based on Good Run List for analyses relying on jet triggers at L1 or HLT (data23_13p6TeV.periodAllYear_DetStatus-v110-pro31-06_MERGED_PHYS_StandardGRL_All_Good_25ns)

***Fig. 16.*** Fraction of good luminosity after DQ criteria are applied with respect to the total ATLAS recorded luminosity.

## 11. SiT Hit Map

A hit map for the SiT detector is shown in Fig. 17 [5] for the first 1.5 M events of run 427929 (LBs 200-206). The distribution of hits in a single SiT plane is shown before and after the signal cleaning for a single track reconstructed per station, a single cluster reconstructed per plane, and only 1 or 2 hits recorded per plane. The "diffractive pattern" is caused by settings of LHC magnet between the ATLAS interaction point and the AFP detectors.

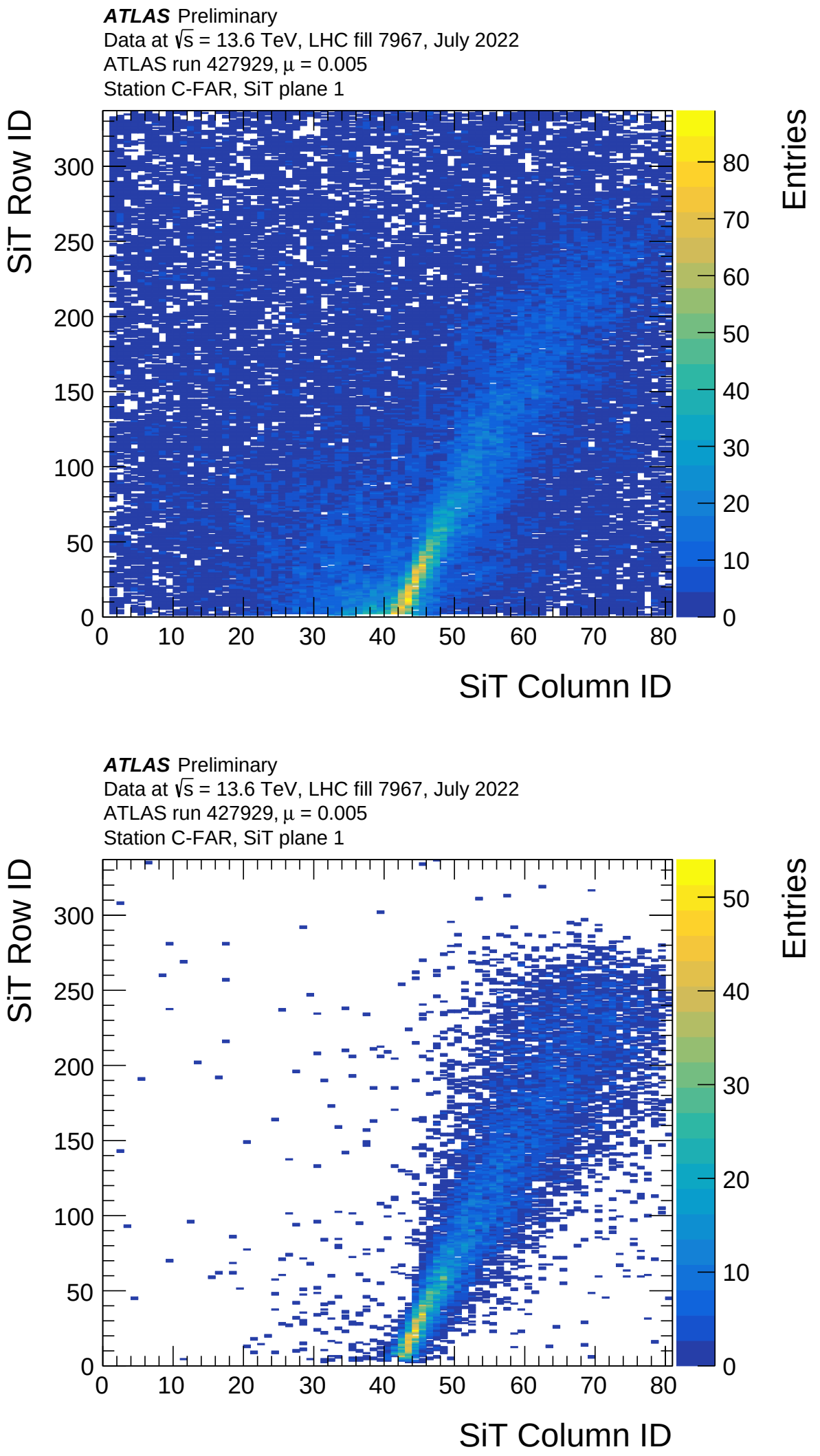

***Fig. 17.*** Distributions of hits in a single SiT plane before and after signal cleaning (from [5]).

## 12. SiT Track Map

The distributions of reconstructed tracks are shown in Fig. 18 [5]. The center of the beam pipe is at position (0, 10 mm). These selection requirements are applied:

- Events triggered by MBTS.
- A reconstructed primary vertex.
- A single track in each station on a given side.
- Expected relation of scattered proton $x$-position in SiT to energy lost $\xi$ AFP in the interaction due to the LHC magnetic field.

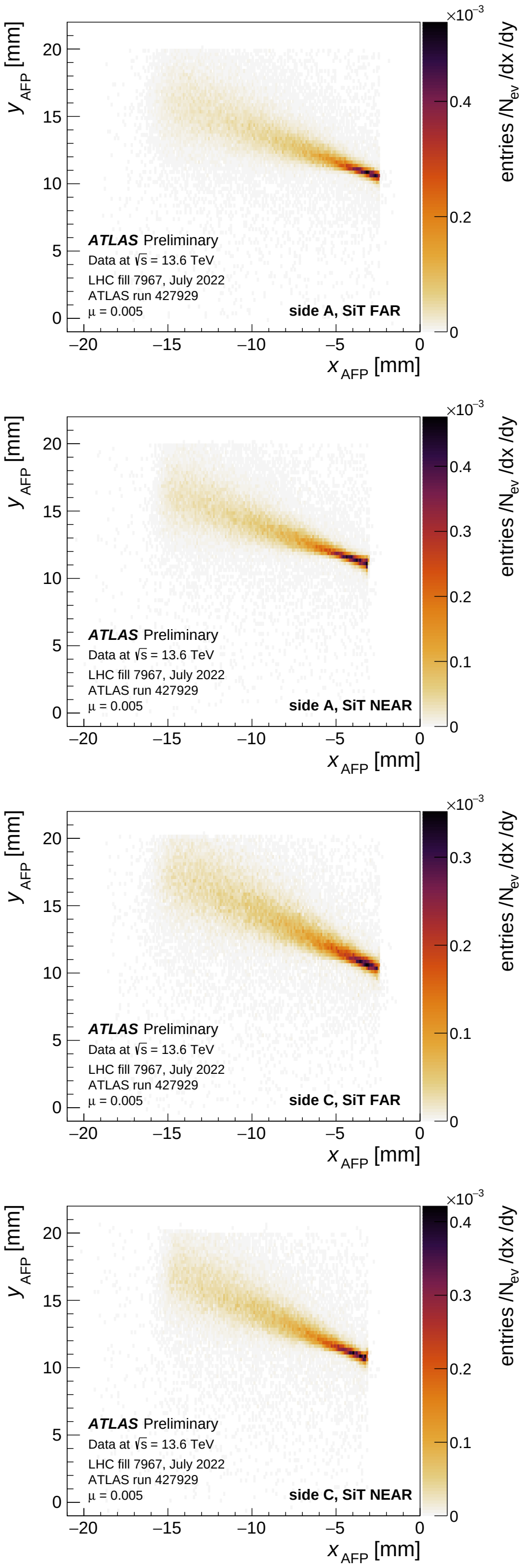

***Fig. 18.*** Distributions of reconstructed tracks for Side A and C, near and far stations (from [5]).

## 13. Correlation AFP and ATLAS Central Detectors

The correlation of the AFP track $x$-position to the charged track multiplicity of the ATLAS Inner Detector (ID) is shown in Fig. 19 [5]. These selection requirements are applied:

- A single AFP track in each station on a given side.
- ID track $p_T > 500$ MeV.
- ID track $|\eta| < 2.5$.
- A reconstructed primary vertex.

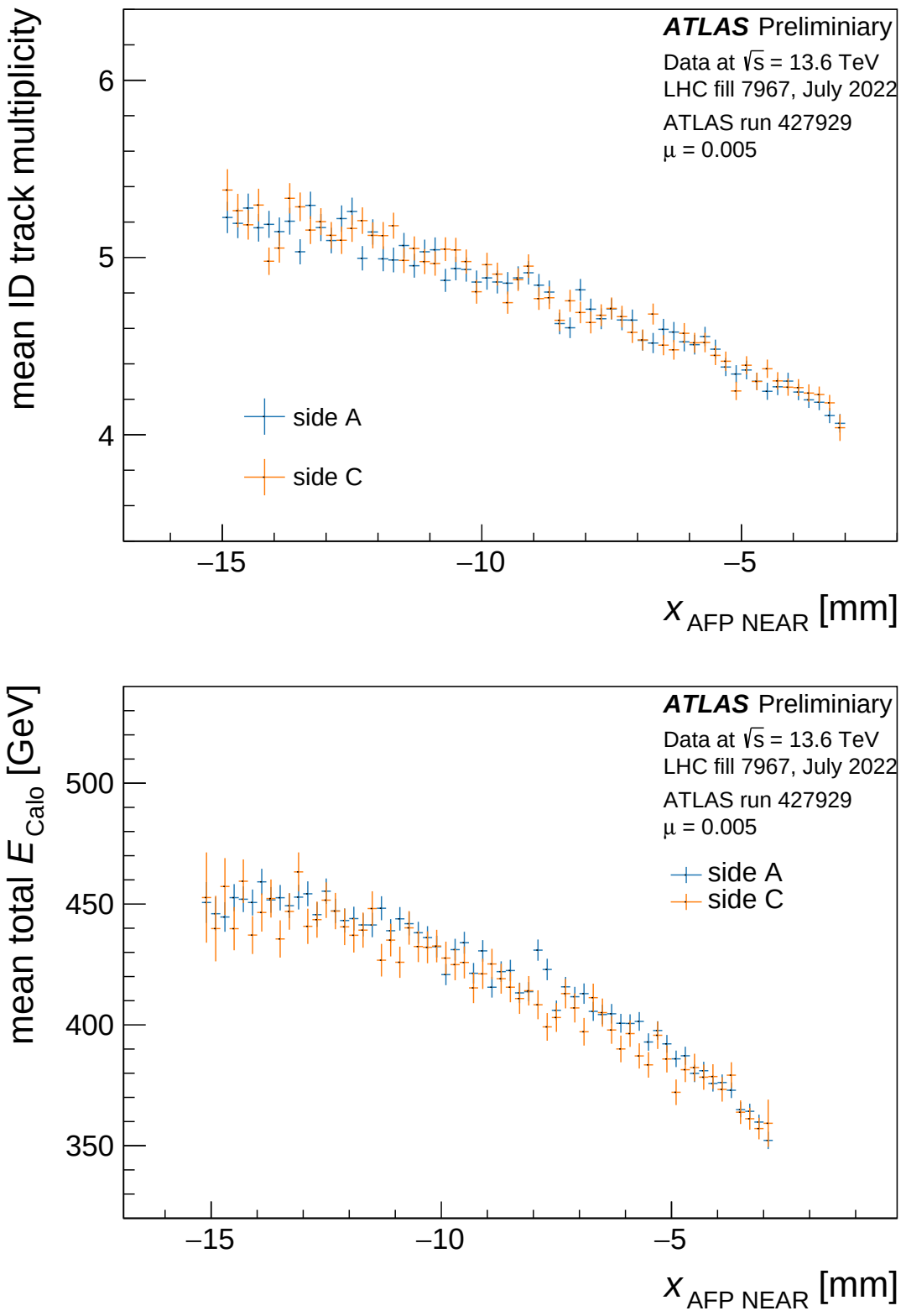

***Fig. 19.*** Correlation of the AFP track $x$-position to the charged track multiplicity of the ATLAS Inner Detector (ID), and correlation of the AFP track $x$-position to the total energy measured by ATLAS calorimeters (from [5]).

Figure 19 [3] also shows the correlation of AFP track $x$-position to the total energy measured by ATLAS calorimeters. The selections are:

- Only one AFP track in each station on a given side.
- A reconstructed primary vertex.

## 14. ToF-SiT Alignment

The correlation of the SiT track $x$-position to the ToF train signal is shown in Fig. 20 [5]. Selection requirements are:

- A single SiT track in the station.
- A single ToF train signal in the station.

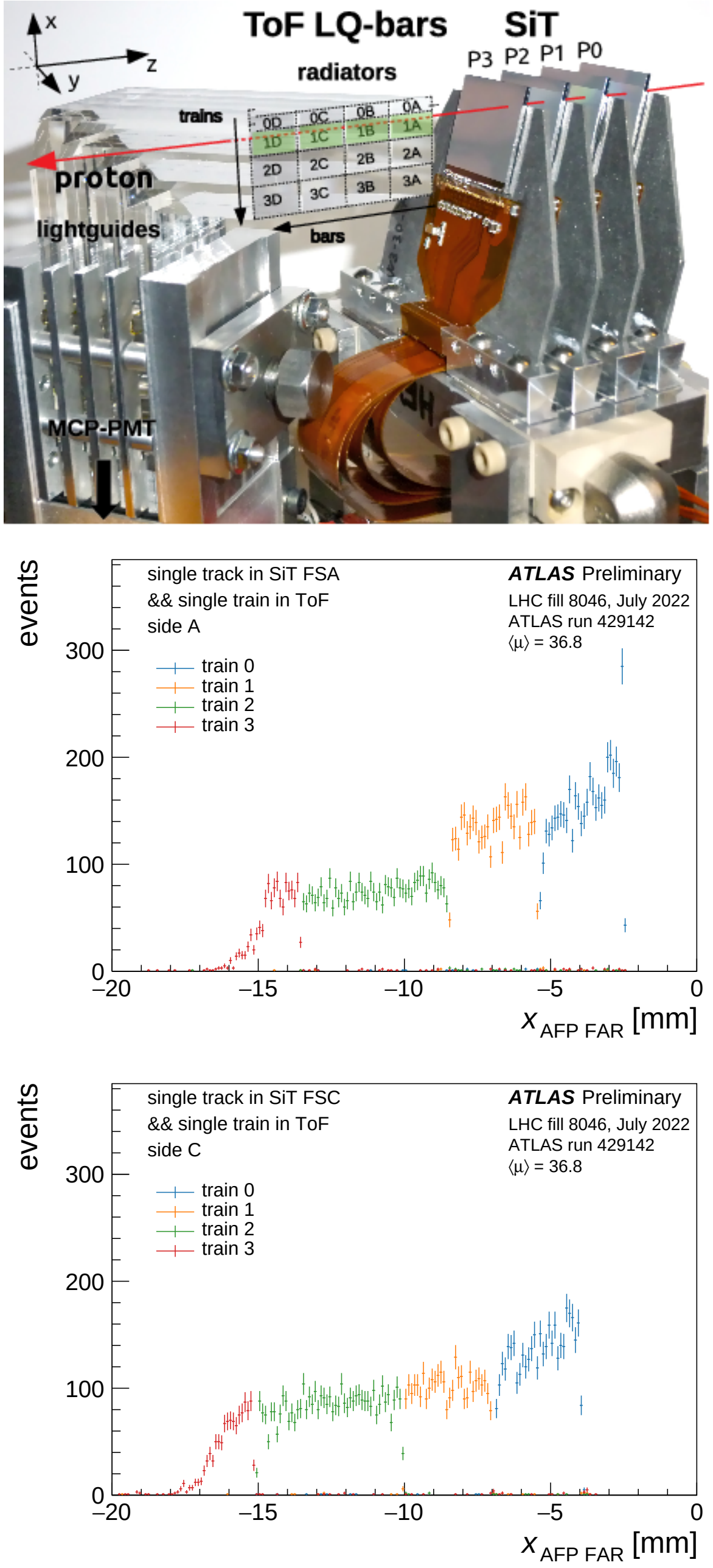

***Fig. 20.*** Correlation of SiT track $x$-position to the ToF train signal (from [5]).

## 15. ToF Efficiency

The probability of observing a hit in the ToF detector during the low-$\mu$ run in July 2022 are shown in Fig. 21 [5]. A Tag-and-Probe method was used, tagged by the SiT for a single track only. A high hit probability is observed in the expected trains.

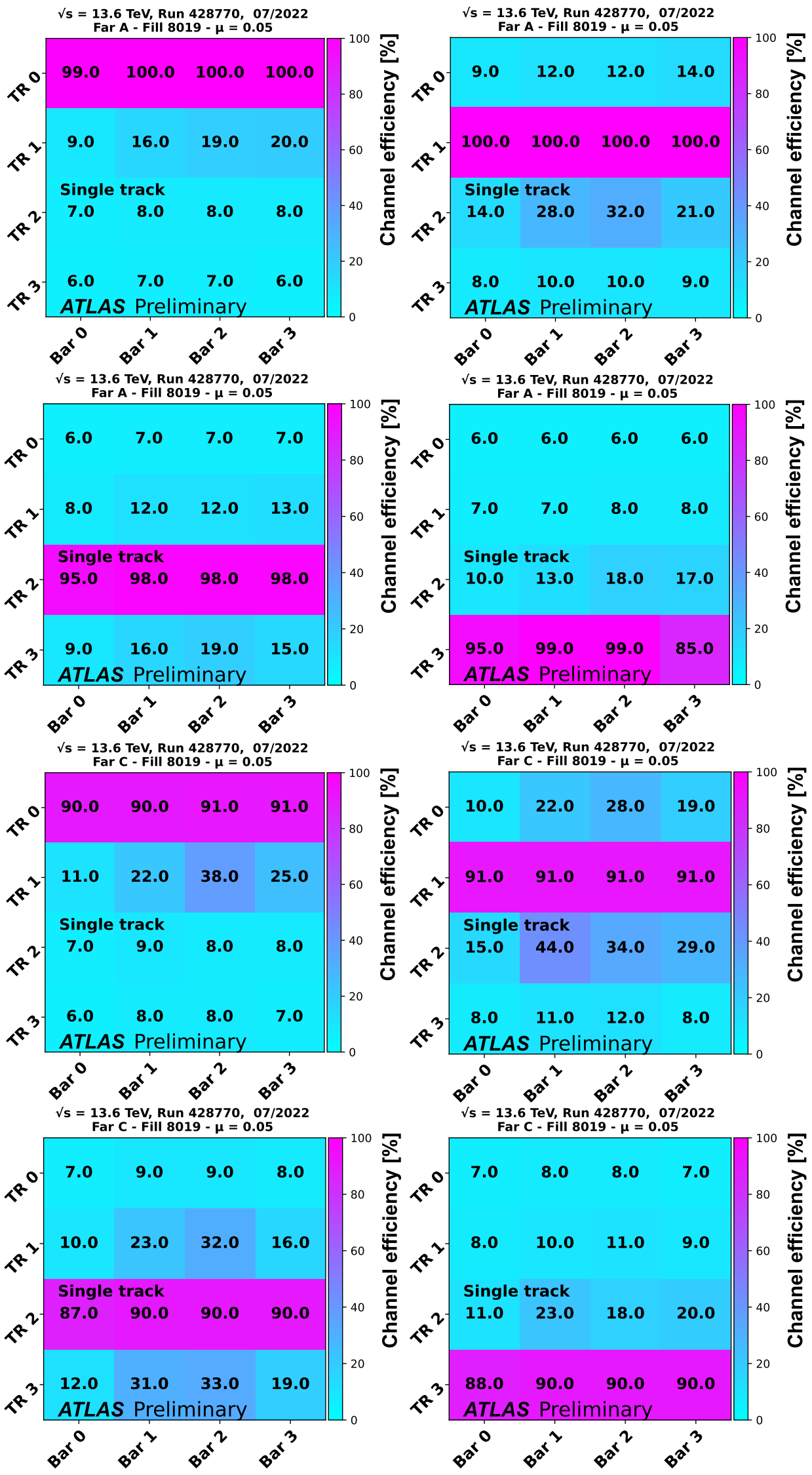

***Fig. 21.*** Probability of observing a hit in the ToF detector during the low-$\mu$ run in July 2022 (from [5]).

## 16. ToF Vertex Matching

The difference between the longitudinal vertex position as measured with AFP ToF and ATLAS Inner Detector (ID) during a $\mu = 0.05$ run taken in July 2022, is shown in Fig. 22 [5]. The resolution is $9.0 \pm 0.1$ mm (30 ps). A small initial background contribution with respect to the signal is observed in low pile-up data-taking conditions.

The advantage of using ToF information is an improvement of the vertex position reconstruction position. Selection requirements are:

- A primary vertex in ATLAS ID.
- A single AFP ToF train signal in each far station.
- Maximum of one hit in each ToF channel.
- A single track in AFP SiT in each far station.
- SiT track position matching the ToF train position.

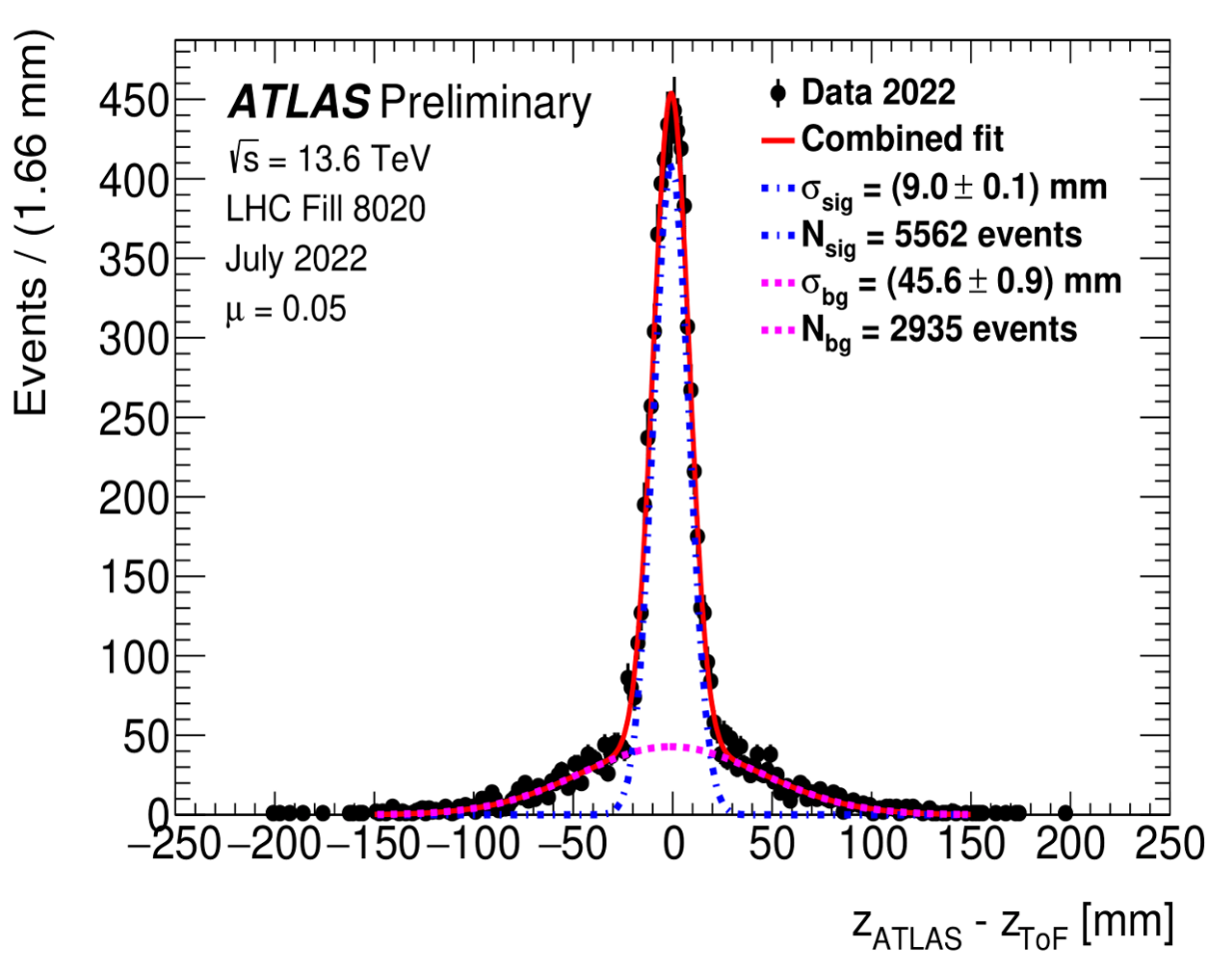

***Fig. 22.*** Difference between longitudinal vertex position measured with AFP ToF and ATLAS Inner Detector (ID) during a $\mu = 0.05$ run taken in July 2022 (from [5]).

## 17. ToF Performance in LHC Run-2

The full-train efficiency was about 4% to 6% as detailed in Fig. 23 [6]. While these low efficiencies were observed, the resolutions of the two ToF detectors measured individually are 21 ps and 28 ps, the vertex reconstruction resolution was $6.0 \pm 2.0$ mm, as shown in Fig. 23 [6] for run 341419 with $\mu = 2$.

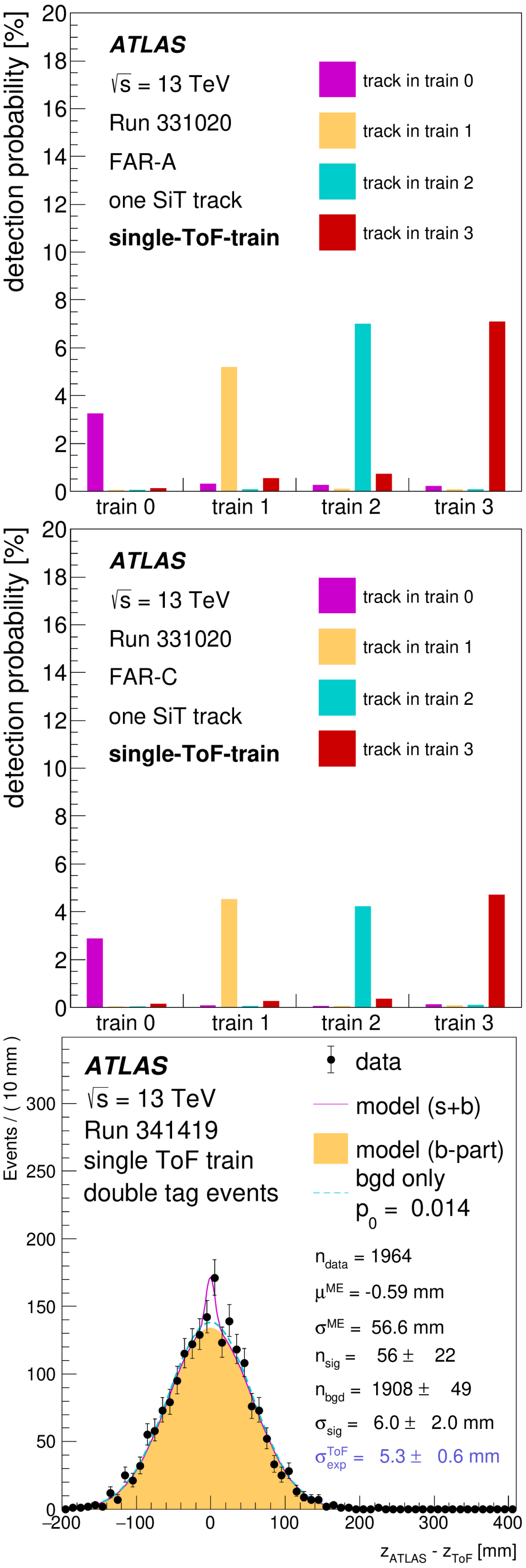

***Fig. 23.*** Full-train efficiency and vertex reconstruction resolution for run 341419 with $\mu = 2$ (from [6]).

## 18. AFP Results

An overview of AFP published results is given in Fig. 24.

**Proton tagging with the one arm AFP detector**
ATL-PHYS-PUB-2017-012 (2017)

**Observation and measurement of forward proton scattering in association with lepton pairs produced via the photon fusion mechanism at ATLAS**
Phys. Rev. Lett. 125 (2020) 261801

**Performance of the ATLAS Forward Proton Time-of-Flight Detector in 2017**
ATL-FWD-PUB-2021-002 (2021)

**Search for an axion-like particle with forward proton scattering in association with photon pairs at ATLAS**
JHEP 2307 (2023) 234

**Performance of the ATLAS Forward Proton Spectrometer during High Luminosity 2017 Data Taking**
ATL-FWD-PUB-2024-001 (2024)

**Performance of the ATLAS forward proton Time-of-Flight detector in Run 2**
JINST 19 (2024) P05054

***Fig. 24.*** Overview of AFP published results.

## 19. Matching of Proton Energy Loss with ATLAS Central Di-Leptons/Di-Photons

The photon-induced di-lepton production with forward proton tag at 13 TeV was studied in the AFP acceptance range $0.02 < \xi < 0.12$, where $\xi$ is the relative proton energy loss [7]. The signal and combinatorial background processes are shown in Fig. 25 [7]. Di-lepton events are studied in the rapidity $y_{\ell\ell}$ versus $m_{\ell\ell}$ plane using 14.6 fb$^{-1}$ [7]. Event are selected with the kinematic matching $|\xi_{\rm AFP} - \xi_{\ell\ell}| < 0.005$ on at least one side (Fig. 26 [7]). Shaded (hatched) areas denote the acceptance (no acceptance) for the AFP stations. Areas neither shaded nor hatched correspond to $\xi$ outside [0,1].

For the light-by-light scattering mediated Axion-Like-Particle (ALP) production, the matching of a photon pair and the proton kinematics is required. Figure 27 [8] shows 441 single matching events in 2017 data. There are no double matching events. The matching requirement is $|\xi_{\rm AFP} - \xi_{\gamma\gamma}| < 0.004 + 0.1\xi_{\gamma\gamma}$.

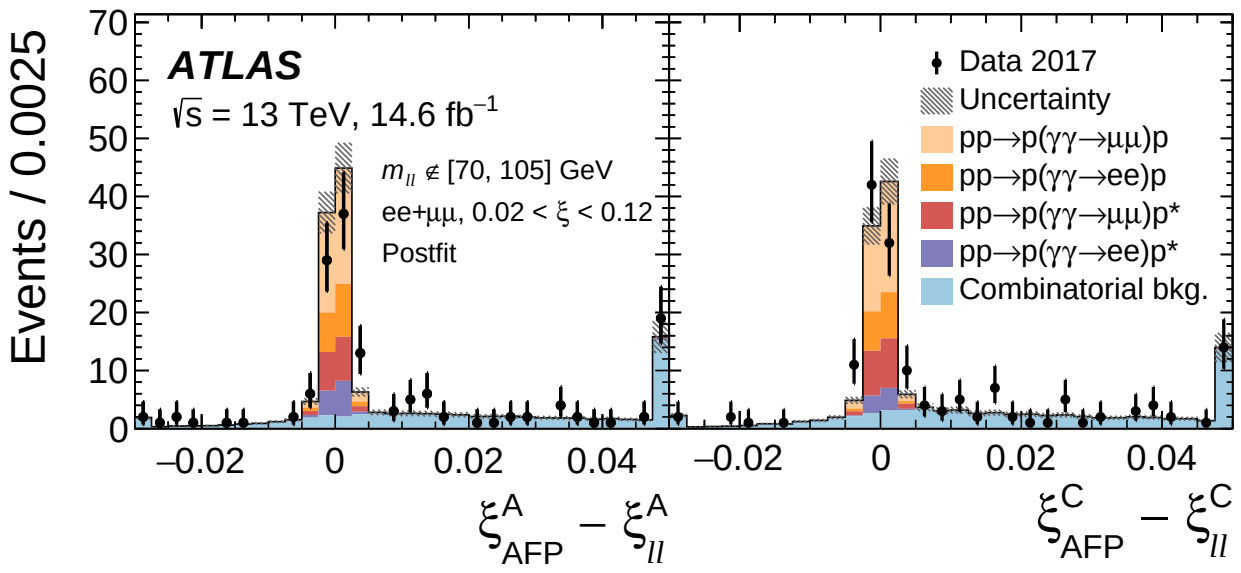

***Fig. 25.*** Di-lepton matching with AFP proton kinematics (from [7]).

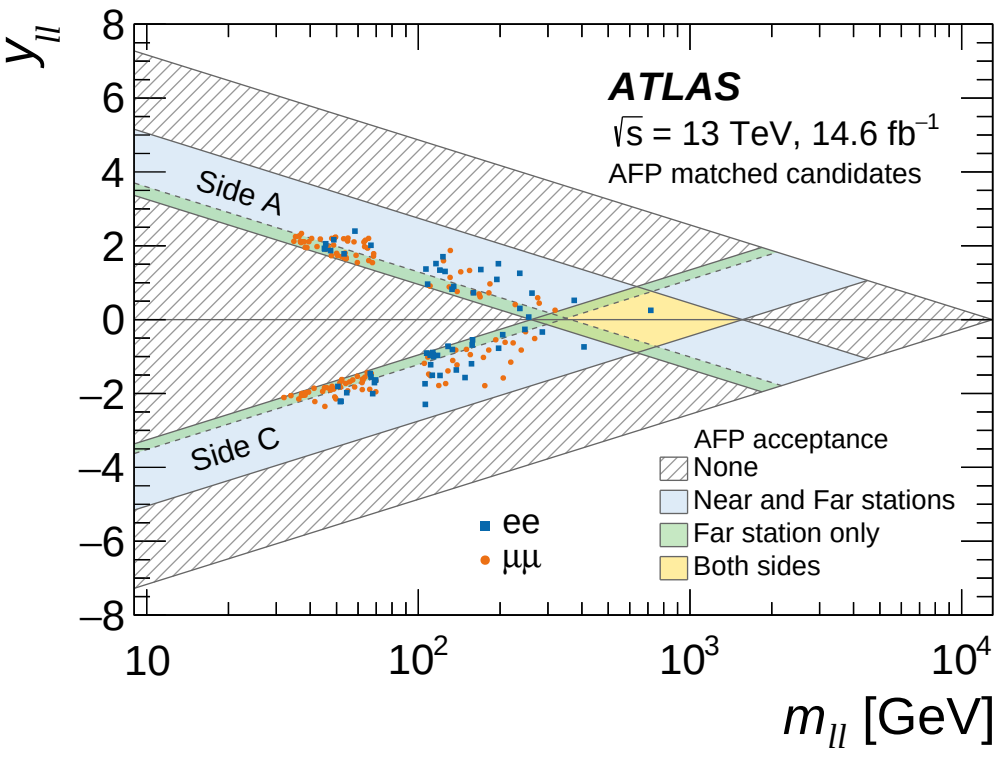

***Fig. 26.*** Di-lepton selected events with AFP tag (from [7]).

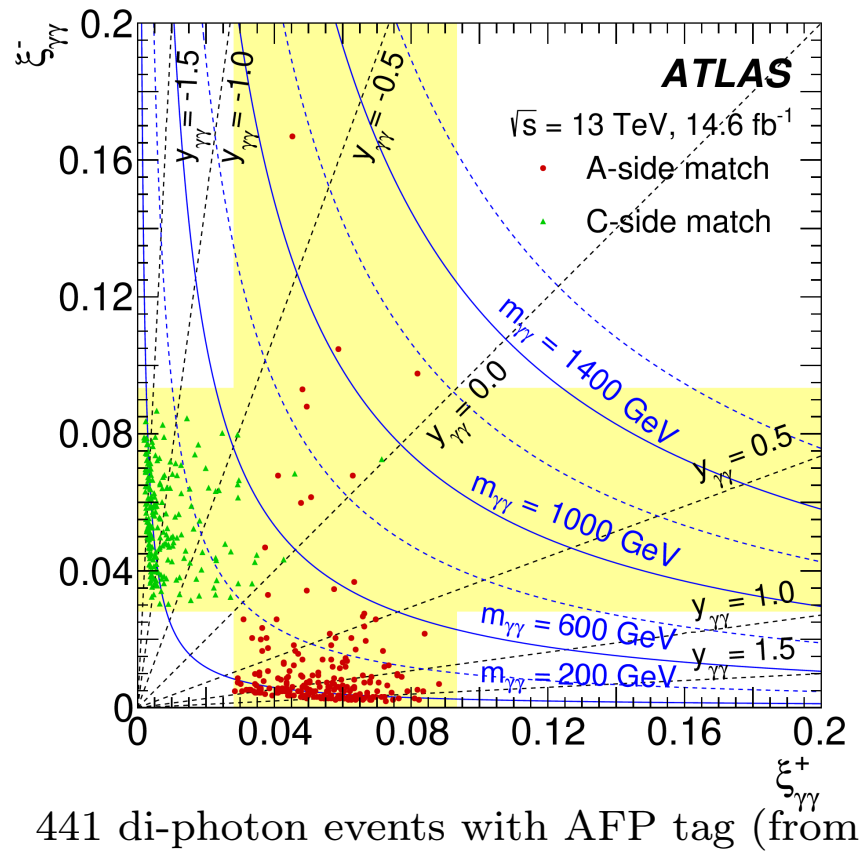

***Fig. 27.*** 441 di-photon events with AFP tag (from [8]).

## 20. Di-Lepton Production in Photon Collisions

For the $\gamma\gamma \to \ell\ell$ analysis, 57 (123) candidates ee + p ($\mu\mu + p$) events are selected [7]. The background-only hypothesis is rejected with a significance $> 5\sigma$ in each channel. Cross-section measurements in the fiducial detector acceptance $\xi \in [0.035, 0.08]$ yield: $\sigma(\text{ee} + \text{p}) = 11.0 \pm 2.6(\text{stat}) \pm 1.2(\text{syst}) \pm 0.3(\text{lumi})$ fb, and $\sigma(\mu\mu + \text{p}) = 7.2 \pm 1.6(\text{stat}) \pm 0.9(\text{syst}) \pm 0.2(\text{lumi})$ fb. A comparison with proton soft survival (no additional soft re-scattering) models gives:
$10.0 \pm 0.8$ fb (ee+p) and $9.4 \pm 0.7$ fb ($\mu\mu$+p) [7].

## 21. Axion-Light-Particle (ALP) Search in Light-by-Light Scattering

The Axion-Like-Particle (ALP) search in the reaction $\gamma\gamma \to \gamma\gamma$ uses an AFP proton tag to reduce the background (Fig. 27) [8], and it leads to limits on the ALP production cross-section and ALP coupling (Figs. 28 [8] and 29 [8]). There is a further rich analysis programme using the AFP detector [5].

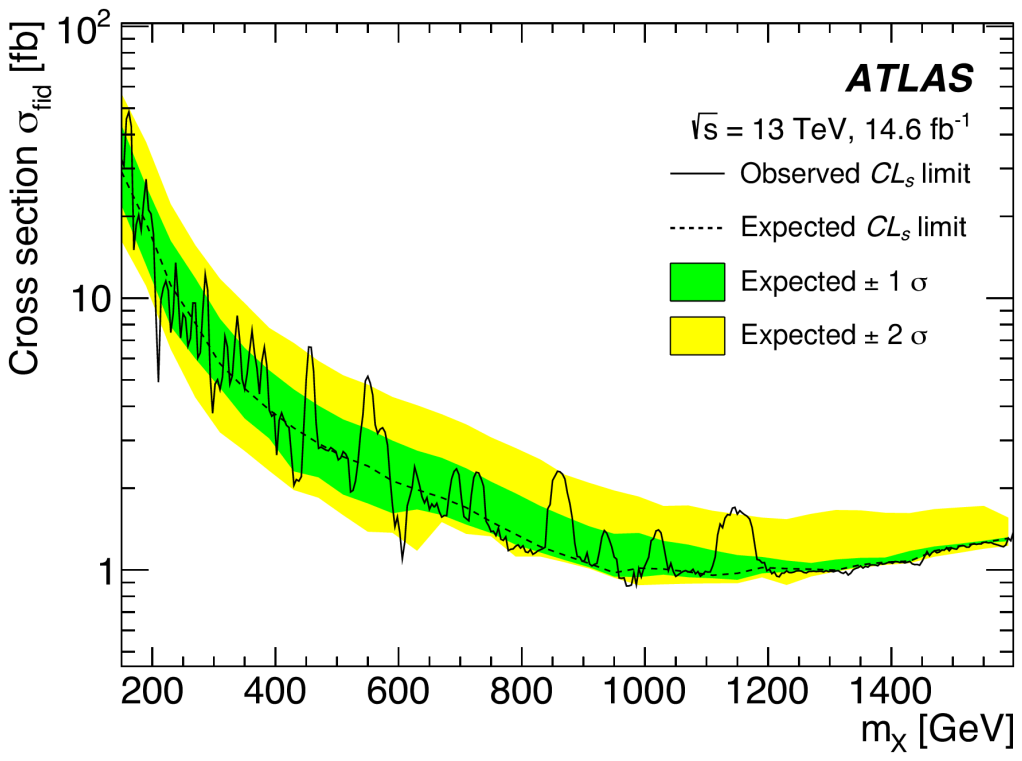

***Fig. 28.*** ALP production cross-section limit (from [8]).
[6]

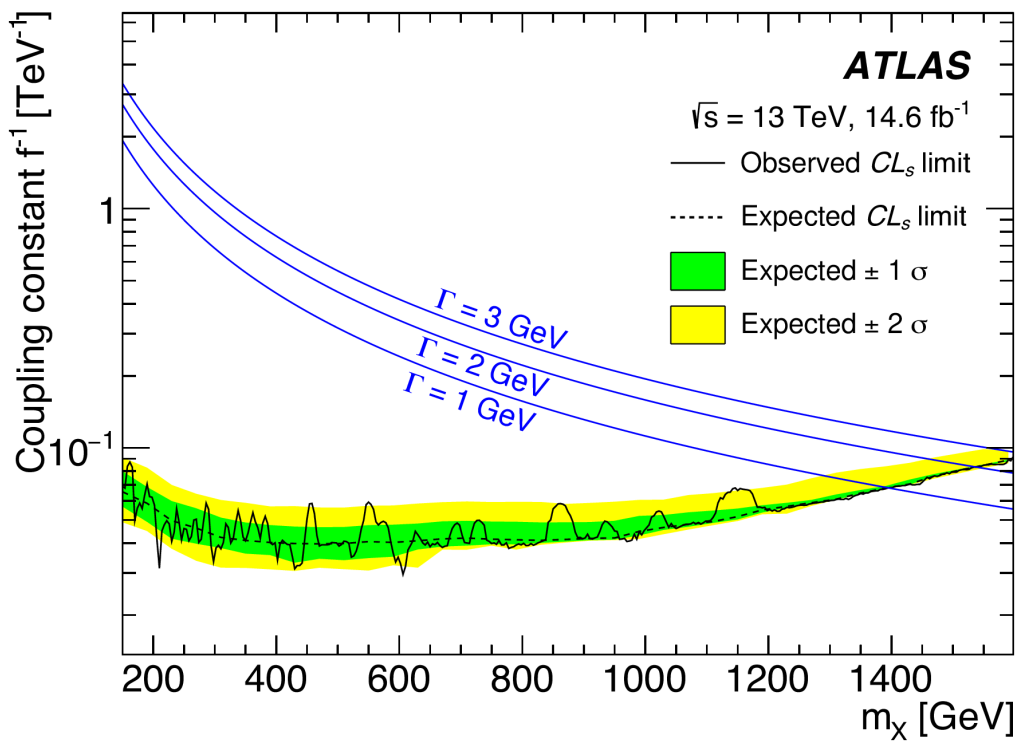

***Fig. 29.*** ALP coupling limit (from [8]).

## 22. Comparison with Previous ALP Results and Extrapolations

The new results have been compared with previous results (Fig. 30 [8]). Extrapolations to LHC Run-3 and HL-LHC luminosities are also shown, based on separating systematic and statistical uncertainties.

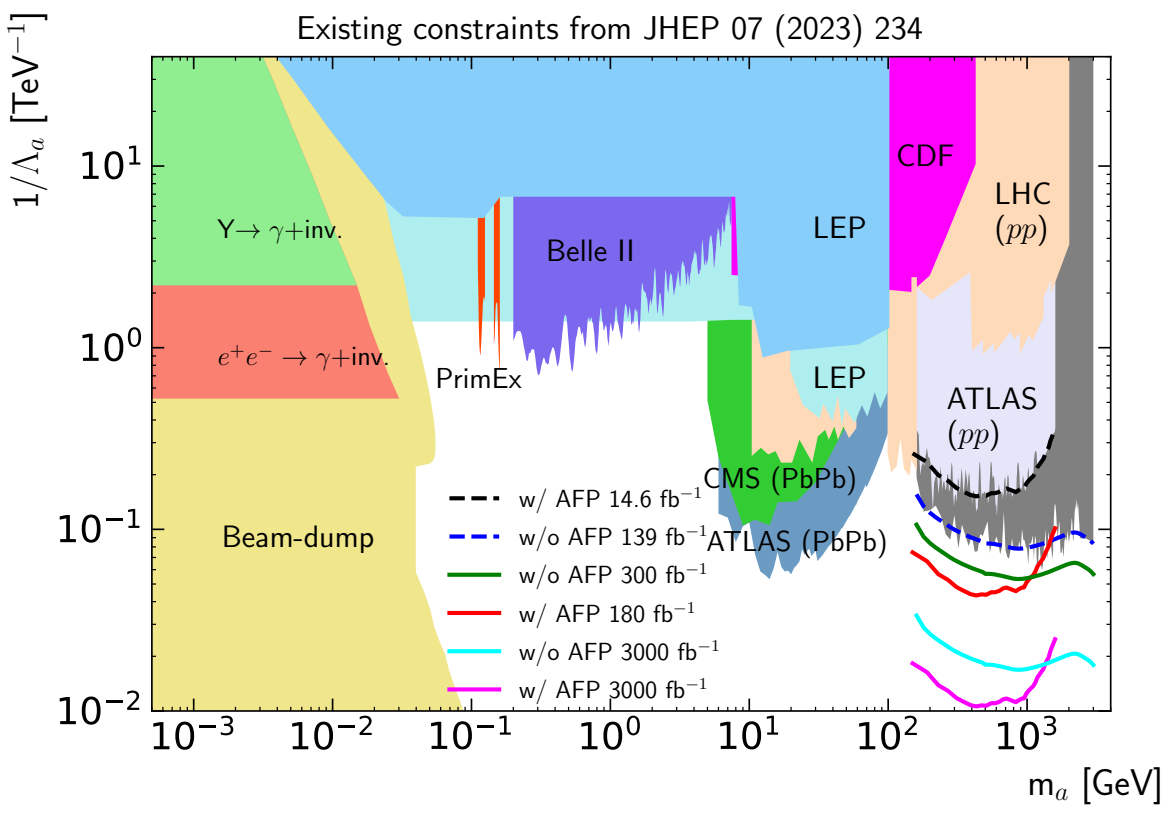

***Fig. 30.*** Overview of AFP results (from [8]) and extrapolations.

## 23. Conclusions

The physics programme with ALFA and AFP is an enhancement of the ATLAS measurement capabilities. The performance of ALFA and AFP-SiT detectors was high. Good efficiency and time reconstruction resolution of AFP-ToF detectors are demonstrated in low-$\mu$ runs. AFP recorded efficiently data during high-$\mu$ runs as well as during special low-$\mu$ runs. Recent ALFA and AFP publications are:

- Measurement of exclusive pion pair production in proton-proton collisions at $\sqrt{s} = 7$ TeV.
- Measurement of total cross-section and $\rho$-parameter from elastic scattering in proton-proton collisions at $\sqrt{s} = 7$ TeV.
- Observation of forward proton scattering in association with lepton pairs in photon fusion in proton-proton collisions at $\sqrt{s} = 13$ TeV.
- ALP with AFP search in Light-by-Light scattering in proton-proton collisions at $\sqrt{s} = 13$ TeV.
- AFP ToF performance in LHC Run-2.

*The research is supported by the Ministry of Education, Youth and Sports of the Czech Republic under the project numbers LTT 17018 and LM 2023040. The author would like to acknowledge the DAAD support under project number 57705645.*

## References

1. ATLAS Collaboration. The ATLAS Experiment at the CERN Large Hadron Collider. *JINST*, 3:S08003, 2008. https://doi.org/10.1088/1748-0221/3/08/S08003.
2. ATLAS Collaboration. Measurement of the total cross section and $\rho$-parameter from elastic scattering in pp collisions at $\sqrt{s} = 13$ TeV with the ATLAS detector. *Eur. Phys. J. C*, 83(5):441, 2023. https://doi.org/10.1140/epjc/s10052-023-11436-8.
3. ATLAS Collaboration. Measurement of exclusive pion pair production in proton–proton collisions at $\sqrt{s} = 7$ TeV with the ATLAS detector. *Eur. Phys. J. C*, 83(7):627, 2023. https://doi.org/10.1140/epjc/s10052-023-11700-x.
4. ATLAS Collaboration. Public ATLAS Luminosity Results for Run-3 of the LHC. https://twiki.cern.ch/twiki/bin/view/AtlasPublic/LuminosityPublicResultsRun3.
5. ATLAS Collaboration. Public Forward Detector Plots for Collision Data. https://twiki.cern.ch/twiki/bin/view/AtlasPublic/ForwardDetPublicResults.
6. ATLAS Collaboration. Performance of the ATLAS forward proton Time-of-Flight detector in Run 2. *JINST*, 19(05):P05054, 2024. https://doi.org/10.1088/1748-0221/19/05/P05054.
7. ATLAS Collaboration. Observation and measurement of forward proton scattering in association with lepton pairs produced via the photon fusion mechanism at ATLAS. *Phys. Rev. Lett.*, 125(26):261801, 2020. https://doi.org/10.1103/PhysRevLett.125.261801.
8. ATLAS Collaboration. Search for an axion-like particle with forward proton scattering in association with photon pairs at ATLAS. *JHEP*, 07:234, 2023. https://doi.org/10.1007/JHEP07(2023)234.